\pdfoutput=1
\documentclass[preprint]{aastex631}
\usepackage{hyperref}
\usepackage{todonotes}
\usepackage{comment}
\usepackage{CJKutf8}

\newcommand\ha{{H$\alpha$}}

\newcommand\kms{\:\rm{\,km\,s^{-1}}}

\newcommand\FLUX{\:{\rm ergs\:cm^{-2}\:s^{-1}}}

\newcommand\eg{{\it e.g.}}

\newcommand\hii{H\,{\footnotesize II}}
\newcommand\sii{[S\,{\footnotesize II}]}
\newcommand\siiL{[S\,{\footnotesize II}] $\lambda\lambda$ 6716,6731}

\newcommand\oiii{[O\,{\footnotesize III}]}

\newcommand\oviL{O\,{\footnotesize VI} $\lambda$ 1032}
\newcommand\ovi{O\,{\footnotesize VI}}
\newcommand\nii{[N\,{\footnotesize II}]}

\newcommand{\EXPU}[3]{\mbox{\rm $#1 \times 10^{#2} \rm\:#3$}}  

\begin{document}

\title{The Dark Energy Camera Magellanic Clouds Emission-Line 
Survey}

\correspondingauthor{Sean D. Points}
\email{sean.points@noirlab.edu}

\author[0000-0002-4596-1337]{Sean D. Points}
\affiliation{NSF's NOIRLab/CTIO\\
Casilla 603 \\
La Serena, Chile}

\author[0000-0002-4134-864X]{Knox S. Long}
\affiliation{Space Telescope Science Institute,
3700 San Martin Dr,
Baltimore MD 21218, USA}
\affiliation{Eureka Scientific, Inc.
2452 Delmer Street, Suite 100,
Oakland, CA 94602-3017, USA}

\author[0000-0003-2379-6518]{William P. Blair}
\affiliation{The William H. Miller III Department of Physics and Astronomy, 
Johns Hopkins University, 3400 N. Charles Street, Baltimore, MD, 21218, USA} 

\author[0009-0006-7371-3484]{Rosa Williams}
\affiliation{Department of Earth and Space Sciences, Columbus State University, 4225 University Avenue, Columbus, Georgia 31907, USA}

\author[0000-0003-3667-574X]{You-Hua Chu \begin{CJK}{UTF8}{bsmi}(朱有花)\end{CJK}}
\affiliation{Department of Physics, National Sun Yet-Sen University, No.\ 70, Lienhai Rd., Kaohsiung 80424, Taiwan}
\affiliation{Institute of Astronomy and Astrophysics, Academia Sinica, No.1, Sec. 4, Roosevelt Rd., Taipei 106216, Taiwan} 

\author[0000-0001-6311-277X]{P. Frank Winkler}
\affiliation{Department of Physics, Middlebury College, Middlebury, VT, 05753, USA}

\author[0000-0001-6311-277X]{Richard L. White}
\affiliation{Space Telescope Science Institute,
3700 San Martin Dr,
Baltimore MD 21218, USA}

\author[0000-0002-4410-5387]{Armin Rest}
\affiliation{Space Telescope Science Institute,
3700 San Martin Dr,
Baltimore MD 21218, USA}
\affiliation{The William H. Miller III Department of Physics and Astronomy, 
Johns Hopkins University, 3400 N. Charles Street, Baltimore, MD, 21218, USA} 

\author[0000-0003-1449-7284]{Chuan-Jui Li \begin{CJK}{UTF8}{bsmi}(李傳睿)\end{CJK}}
\affiliation{Institute of Astronomy and Astrophysics, Academia Sinica, No.1, Sec. 4, Roosevelt Rd., Taipei 106216, Taiwan} 



\author[0000-0001-5567-1301]{Francisco Valdes}
\affiliation{NSF's NOIRLab, 950  Cherry Ave, Tucson, AZ 85719}

\begin{abstract}

We have used the Dark Energy Camera (DECam) on the CTIO Blanco 4-m 
telescope to perform a new emission-line survey of the Large Magellanic
Cloud (LMC) using narrow-band \ha\
and \sii\ filters in addition to a continuum band for use in creating pure emission-line images.  We refer to this new survey as DeMCELS, to distinguish it from the earlier Magellanic Cloud Emission Line Survey (MCELS) that has been in service for nearly 30 years. DeMCELS covers $\rm \sim 54~ degrees^{2}$, encompassing most of the bright optical disk of the LMC.  With DECam’s pixel size of only 0\farcs27, our DeMCELS survey
provides a seeing-limited improvement of $3$ – $5 \times$ over MCELS and is comparable in depth, with surface brightness limits of \EXPU{3.3}{-17}{\FLUX~arcsec^{-2}} and \EXPU{2.9}{-17}{\FLUX~arcsec^{-2}} in \ha\ and \sii, respectively.
DeMCELS provides detailed morphological information on nebulae of 
all scales, from the largest supershells to individual \hii\ regions and 
supernova remnants, to bubbles of emission surrounding individual stars, 
and even to faint structures in the diffuse ionized gas of the LMC. Many 
complex regions of emission show significant variations in the ratio of 
\sii\ to \ha, a sign of the mixture of shocks from stellar winds and/or 
supernovae with photoionization by embedded hot, young stars. We present 
the details of the observing strategy and data processing for this survey, 
and show selected results in comparison with previous data. A companion project for the Small Magellanic Cloud is in progress and will be reported separately.  We are making these new data available to the community at large via the NOIRLab's Data Lab site.

\end{abstract}


\section{Introduction} \label{sec:intro}

The Large Magellanic Cloud (LMC) is the largest and most massive satellite
galaxy of the Milky Way.  At a distance of $49.59\pm0.09$~kpc
\citep{Pietrzynski19}, and seen along a line of sight with low Galactic 
foreground absorption, the LMC is the best laboratory for studying a wide
variety of astrophysical phenomena, including the life cycle of stars (star 
formation and stellar death in all forms), the interstellar medium (ISM), 
and the interplay between them.  With a mass of $1.8 \times 
10^{11}~M_{\odot}$ \citep{Shipp_2021}, the LMC has a star formation rate of
$0.2~M_{\odot}$~yr$^{-1}$ \citep{HZ2009}, dominated by the spectacular 
30~Dor complex, and has a metallicity of $[{\rm Fe}/{\rm H}] =
-0.42$~dex \citep{Choudhury2021}, similar to that of M33.  The LMC contains
complexes of emission-line gas, excited by both photoionization from young, 
hot stars, and indirectly by shocks from stellar winds and the supernovae
that have exploded over time.  Even away from regions of active star formation, 
little-studied faint diffuse gas fills the interstellar regions of the galaxy, 
ionized by starlight leaking out of the star forming regions.

At optical wavelengths, the ISM is observed primarily in the Balmer lines of
hydrogen, and various forbidden lines of intermediate mass elements like O,
N, and S.  The principal method of distinguishing shocked from photoionized
gas has traditionally been the ratio of \siiL\ to \ha\ \citep[][and 
references therein]{mathewson83}.  Images
of portions of the LMC have been accumulated by many observers over the
decades, but in the CCD era, the most widely used global emission-line survey of
both the Large and Small Magellanic Clouds has been the Magellanic Clouds
Emission Line Survey \citep[MCELS,][]{smith99}.  From 1997--2001, MCELS
produced data covering the central 8\arcdeg\ $\times$ 8\arcdeg\ of the LMC 
using the University of Michigan/CTIO Curtis Schmidt telescope with the Newtonian 
focus CCD and five filters: three centered on emission lines (\oiii, \ha,
and \sii) and two continuum bands (green and red) for stars and continuum 
subtraction.  The individual MCELS observations covered a 1\fdg4 $\times$
1\fdg4 field of view with a pixel scale of 2\farcs3~pixel$^{-1}$ and an
effective spatial resolution of $\sim$5\arcsec.

The MCELS data have been an extremely powerful tool for investigating many
aspects of the interstellar component of the LMC for more than 20 
years.  Because the ratio of \sii\ to \ha\ is, to first order, a diagnostic of shocks vs.
photoionization\footnote{An updated and more nuanced 
discussion of the \sii/\ha\ ratio is provided in Sec. 6 of \citet{long22}.}, 
MCELS has provided critical data in
studies of supernova remnants \citep[SNRs,][]{Kavanagh15a, Kavanagh15b, Kavanagh22, Yew21, 
deHorta12, Crawford10}, superbubbles, and supergiant shells 
\citep{dunne01, Warth14, Kavanagh20, Collischon21, Sasaki22}.   MCELS data 
have also been used in a variety of investigations such as: (1) determining the 
optical depth of \hii\ regions \citep{Pellegrini12}; (2) searching for 
planetary nebulae \citep{Reid13}; (3) investigating the physical conditions
of Wolf-Rayet nebulae \citep{Hung21}; (4) helping to separate the thermal
and non-thermal radio emission in the LMC \citep{Hassani22}; and many other
multiwavelength campaigns in the LMC \citep[][to mention a few]{Kim03, blair06, Maggi16, bozzetto17}. 

Although MCELS was revolutionary for its time, advances in instrument
technology and computing power are now allowing larger aperture telescopes
the ability to image wide fields of view at significantly higher angular
resolution. In this paper, we have undertaken the next generation CCD
emission-line survey of the LMC using the Dark Energy Camera \citep[DECam;]
[]{Honscheid08, Flaugher15} on the CTIO Blanco 4-m telescope, starting with
imagery at \ha\ and \sii, and a red continuum band. 
Our DECam narrow-band imaging survey 
provides a significant improvement in spatial resolution ($\ge3\times$, depending on seeing) 
and reaches comparable depth to the MCELS data, thus  greatly improving 
investigations of the detailed morphology of both bright and faint, diffuse ISM structures in the MCs on a global scale.  A companion survey covering the Small Magellanic Cloud with five DECam fields is in progress and will be reported separately.

The \ha\ and \sii\ observations described below were not obtained at the same
time. In this paper, we combine the DECam N662 (\ha) images of the LMC from a
survey led by Puzia (PropID: 2018A-0909; 2018B-0908) and a separate N673
(\sii) and DES r$^\prime$ surveyled by Points (PropID: 2021B-0060) into a
single unified analysis.  We describe the observations in \S\ref{sec:obs} and
the data reduction and quality assessments in \S\ref{sec:redx_qual}. Additional
data processing beyond the initial pipeline reductions is discussed in
\S\ref{sec:kred}.  In \S\ref{sec:results}, we provide some examples and
comparisons for various types of objects between our survey and MCELS to
provide a sense of the power of this new higher resolution emission-line
resource. We are making the processed data available to the community at large
via the NOIRLab DataLab site, as described in \S\ref{sec:datalab}.

\section{Observations}\label{sec:obs} 

Our DECam Magellanic Cloud Emission-Line Survey (hereinafter DeMCELS) consists
of 20 overlapping and dithered DECam pointings in several filters covering the
entire bright optical disk of the LMC, approximately 54~deg$^2$.  Each DECam
image covers a roughly circular region of 1\fdg8 diameter with an array of 60
2048 $\times$ 4096 CCDs and a pixel size of 0\farcs27.  Hence, depending on
seeing conditions, a spatial resolution improvement of a factor of 3 to 5 or
more can be expected over MCELS.  

\begin{figure} [ht!]
\begin{center}
\includegraphics[scale=0.70]{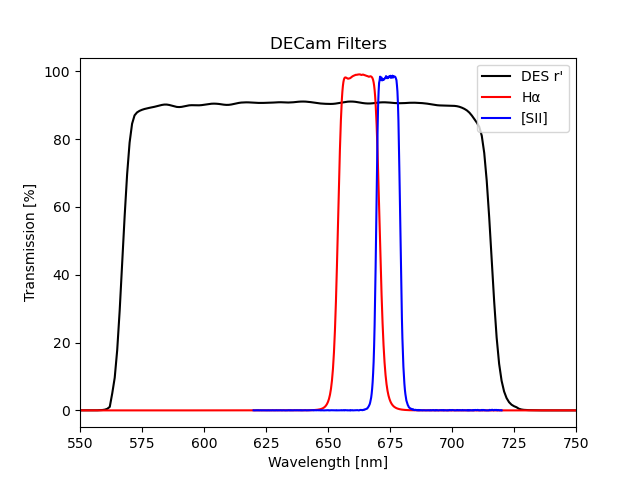}
\caption{\label{filters}DECam filter curves used in this 
investigation with DES r$^{\prime}$ (black), N662 (red), and N673 (blue).}  
\end{center}
\end{figure}

Our observations used the following filters: the narrow-band N662 (\ha  
 + \nii\ $\lambda\lambda 6548,6583$) and N673 (\sii\ $\lambda\lambda 6716,6731$)
to image optical emission lines, and the DES r$^{\prime}$ filter which was used
for continuum subtraction.  The properties of these filters are listed in
Table~1 and shown in Figure~\ref{filters}.  We note that even though the N662
filter contains emission from both \ha\ and \nii\ $\lambda\lambda 6548,6583$,
we will refer to it as the \ha\ filter throughout this work.  This is because
the \nii\ emission lines are generally weak due to the lower N abundance in the
LMC and because the spatial distribution of both \ha\ and \nii\ is expected to
be quite similar.\footnote{From the compilations of \hii\ region spectra by
\citet{Dufour75} and \citet{CarlosReyes15}, we find the \nii\
$\lambda6583$/\ha\ ratio to be 0.05$\pm$0.02 for the LMC.}  The N673 filter
will be referred to as the \sii\ filter.  

\begin{deluxetable}{lcc} 
\tablecaption{DECam Filter Properties}
\tablehead{
  \colhead{Filter} &
  \colhead{Central Wavelength} &
  \colhead{FWHM}
\\
  \colhead{} &
  \colhead{(\AA)} &
  \colhead{(\AA)}
}
\tabletypesize{\scriptsize}
\tablewidth{0pt}
\startdata
DES r$^{\prime}$ & 6420 & 1480 \\
N662 (\ha)       & 6620 & 160  \\
N673 ({[S\,II]}) & 6730 & 100  \\
\enddata
\label{tab:filters}
\end{deluxetable}

The total desired exposure time per field in \ha\ was 4980~s, chosen 
to reach a surface brightness sensitivity limit of $\sim 3 \times 
10^{-18}$~erg~s$^{-1}$~cm$^{-2}$~arcsec$^{-2}$.  This was normally divided 
into six short exposures of 30~s each and six long exposures of 800~s 
for each field.  The telescope was dithered between each of the individual
exposures for each field, allowing the gaps between the DECam individual
detectors to be filled in.  The total exposure time per field for the \sii\
observations was double the exposure time of the \ha\ observations, in
order to reach a similar surface brightness limit because of the fainter 
S emission lines.  These observations were normally divided into 12 short
exposures of 60~s each and 12 long exposures of 800~s, although some fields
with bright emission and/or heavy crowding were instead observed with 24
long exposures of 400~s.  The exposure times of the r$^\prime$ observations
were selected to reach the expected photometric depth of point sources in the 
narrow-band observations in order to facilitate continuum subtraction. 
This corresponds to short and long exposure times of 8~s and 60~s, 
respectively.

\begin{figure} [ht]
\begin{center}
\includegraphics[scale=0.60]{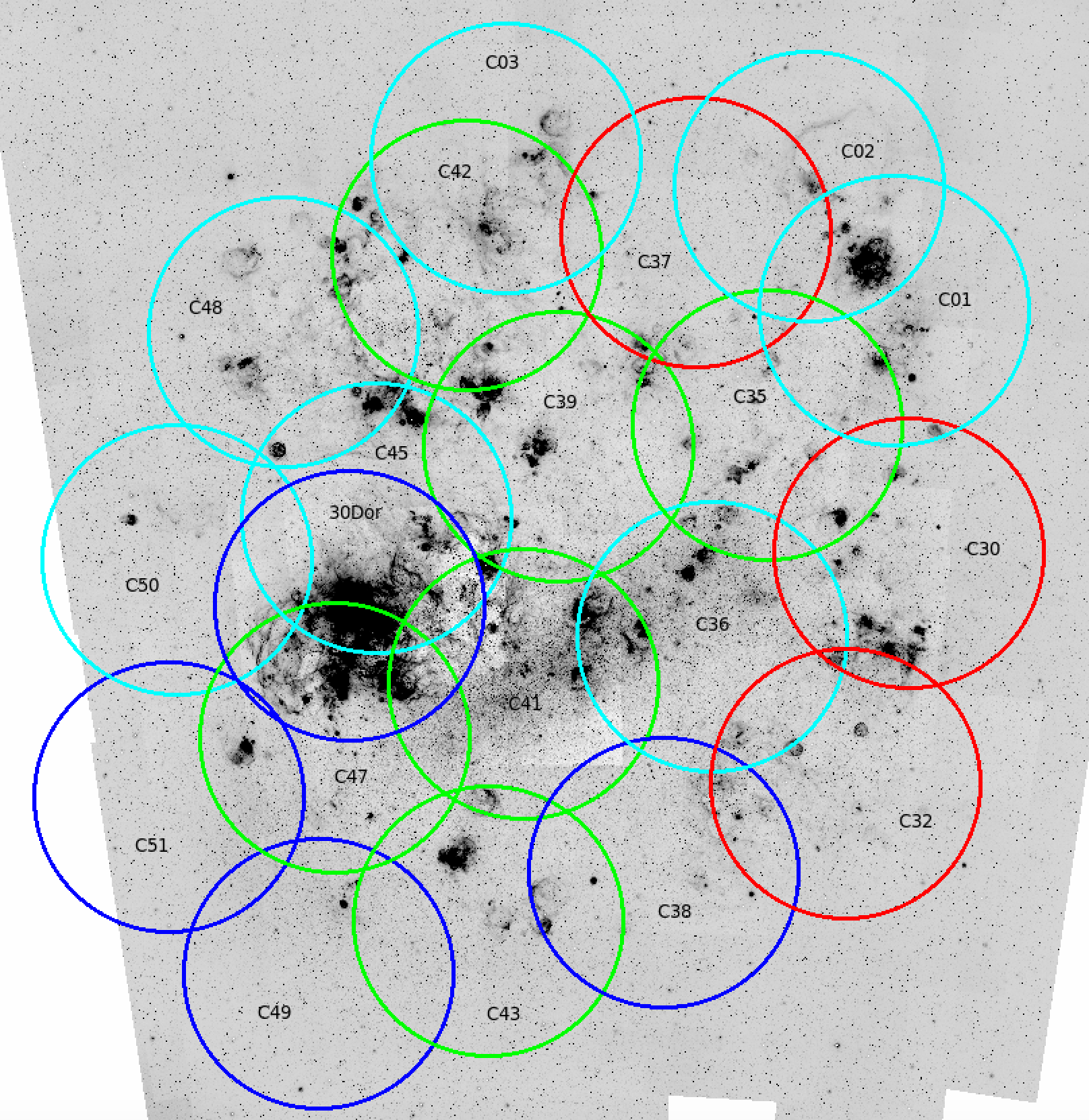}
\caption{\label{lmc-overview}  MCELS \ha\ image of the LMC with 2\arcdeg\
diameter circles representing the approximate DECam footprint overlaid for all
narrow-band data of the LMC found in the NOIRLab Astro Data Archive.  As 
discussed in 
more detail in \S\ref{sec:redx_qual}, colors of the
circles represent the completeness of the long observations compared to a
desired uniform depth: $\geq 2/3$ complete in \ha\ and \sii\ (green); $\geq
2/3$ complete in \ha\ and $\leq 2/3$ complete in \sii\ (blue); $\leq 2/3$
complete in \ha\ and $\geq 2/3$ complete in \sii\ (cyan); and $\leq 2/3$
complete in both \ha\ and \sii\ (red).}
\end{center}
\end{figure}

A summary of all of the DECam observations for each field per filter
is presented in Table~2. Inspection of this Table shows that, given observing conditions and the time available, we were not able to observe all
of the fields to the desired depth, with some fields lacking some or all observations in
either \ha\ or \sii.
We present the MCELS \ha\ image of the LMC with the DECam fields overlaid in Figure~\ref{lmc-overview}.  The color coding in this Figure shows where data shortcomings are present.

\begin{deluxetable}{lccccc}
\tablecaption{DECam Observations}
\tablehead{
 \colhead{Field ID} & 
 \colhead{R.A. (J2000)} & 
 \colhead{Decl. (J2000)} & 
 \multicolumn{3}{c}{Filters}
\\
 \colhead{} & 
 \colhead{} & 
 \colhead{} & 
 \colhead{DES r$^\prime$} &
 \colhead{N662}  &
 \colhead{N673}  
\\
 \colhead{} & 
 \colhead{(deg)} & 
 \colhead{(deg)} & 
 \colhead{(s)} & 
 \colhead{(s)} & 
 \colhead{(s)}  
}
\tabletypesize{\scriptsize}
\tablewidth{0pt}
\startdata
LMC\_30Dor & 84.640830 & $-$69.085860 &  3 $\times$ 20.0   &  4 $\times$ 30.0  & \nodata           \\
           &           &              &  3 $\times$ 120.0  &  4 $\times$ 300.0 & \nodata           \\
LMC\_c01   & 73.591856 & $-$66.807029 & 12 $\times$ 8.0    &  5 $\times$ 30.0  & 12 $\times$ 60.0  \\
           &           &              & 12 $\times$ 60.0   &  \nodata          & 12 $\times$ 800.0 \\
LMC\_c02   & 75.395313 & $-$65.915355 & 14 $\times$ 8.0    &  \nodata          & 15 $\times$ 60.0  \\
           &           &              & 12 $\times$ 60.0   &  \nodata          & 12 $\times$ 800.0 \\
LMC\_c03   & 81.041359 & $-$65.749791 & 14 $\times$ 8.0    &  \nodata          & 14 $\times$ 60.0  \\
           &           &              & 12 $\times$ 60.0   &  \nodata          & 12 $\times$ 800.0 \\
LMC\_c30   & 72.804846 & $-$68.628810 & 16 $\times$ 8.0    &  6 $\times$ 30.0  & 13 $\times$ 60.0  \\
           &           &              & 14 $\times$ 60.0   &  7 $\times$ 800.0 & 12 $\times$ 800.0 \\
LMC\_c32   & 73.662449 & $-$70.431310 & 12 $\times$ 8.0    &  5 $\times$ 30.0  & 12 $\times$ 60.0  \\
           &           &              & 12 $\times$ 60.0   &  6 $\times$ 800.0 & 12 $\times$ 800.0 \\
LMC\_c35   & 75.925342 & $-$67.750762 & 12 $\times$ 8.0    &  4 $\times$ 30.0  & 12 $\times$ 60.0  \\
           &           &              & 12 $\times$ 60.0   &  5 $\times$ 800.0 & 12 $\times$ 800.0 \\
LMC\_c36   & 76.827761 & $-$69.387311 & 12 $\times$ 8.0    &  5 $\times$ 30.0  & 12 $\times$ 60.0  \\
           &           &              & 24 $\times$ 30.0   &  6 $\times$ 800.0 & 24 $\times$ 400.0 \\
LMC\_c37   & 77.475267 & $-$66.306975 & 12 $\times$ 8.0    &  4 $\times$ 30.0  & 12 $\times$ 60.0  \\
           &           &              &  2 $\times$ 60.0   &  \nodata          &  2 $\times$ 800.0 \\
LMC\_c38   & 77.761190 & $-$71.196145 & 12 $\times$ 8.0    &  4 $\times$ 30.0  & 12 $\times$ 60.0  \\
           &           &              & 12 $\times$ 60.0   &  6 $\times$ 800.0 & 12 $\times$ 800.0 \\
LMC\_c39   & 80.141220 & $-$67.952984 & 12 $\times$ 8.0    &  4 $\times$ 30.0  & 12 $\times$ 60.0  \\
           &           &              & 12 $\times$ 60.0   &  5 $\times$ 800.0 & 12 $\times$ 800.0 \\
LMC\_c41   & 80.971450 & $-$69.756543 & 12 $\times$ 8.0    &  4 $\times$ 30.0  & 12 $\times$ 60.0  \\
           &           &              & 28 $\times$ 30.0   &  6 $\times$ 800.0 & 24 $\times$ 400.0 \\
LMC\_c42   & 81.831476 & $-$66.471265 & 12 $\times$ 8.0    &  5 $\times$ 30.0  & 12 $\times$ 60.0  \\
           &           &              & 12 $\times$ 60.0   &  6 $\times$ 800.0 & 12 $\times$ 800.0 \\
LMC\_c43   & 81.923156 & $-$71.556051 & 11 $\times$ 8.0    &  4 $\times$ 30.0  & 12 $\times$ 60.0  \\
           &           &              & 11 $\times$ 60.0   &  6 $\times$ 800.0 & 12 $\times$ 800.0 \\
LMC\_c45   & 83.928142 & $-$68.444092 & 12 $\times$ 8.0    &  5 $\times$ 30.0  & 12 $\times$ 60.0  \\
           &           &              & 24 $\times$ 30.0   &  6 $\times$ 800.0 & 24 $\times$ 400.0 \\
LMC\_c47   & 85.215270 & $-$70.080852 & 12 $\times$ 8.0    &  4 $\times$ 30.0  & 12 $\times$ 60.0  \\
           &           &              & 28 $\times$ 30.0   &  6 $\times$ 800.0 & 27 $\times$ 400.0 \\
LMC\_c48   & 85.481748 & $-$66.961984 & 11 $\times$ 8.0    &  4 $\times$ 30.0  & 11 $\times$ 60.0  \\
           &           &              & 11 $\times$ 60.0   &  7 $\times$ 800.0 & 11 $\times$ 800.0 \\
LMC\_c49   & 86.134066 & $-$71.855063 & \nodata            &  4 $\times$ 30.0  &  \nodata          \\
           &           &              & \nodata            &  6 $\times$ 800.0 &  \nodata          \\
LMC\_c50   & 88.158561 & $-$68.600023 & 11 $\times$ 8.0    &  4 $\times$ 30.0  & 11 $\times$ 60.0  \\
           &           &              & 11 $\times$ 60.0   &  7 $\times$ 800.0 & 11 $\times$ 800.0 \\
LMC\_c51   & 89.083554 & $-$70.376535 & \nodata            &  4 $\times$ 30.0  &  \nodata          \\
           &           &              & \nodata            &  6 $\times$ 800.0 &  \nodata          \\
\enddata
\label{tab:observations}
\end{deluxetable}

\section{Data Reduction and Quality Assessment}\label{sec:redx_qual}

We begin with the object images, processed with the DECam Community Pipeline 
\cite[hereafter the DCP,][]{DECamCP} for bias subtraction and flat-fielding
(i.e., ooi images).  
The DCP produces flux-calibrated images with accurate astrometric solutions based on 
Gaia data.  One portion of the Pipeline process is to remove sky background and a 
pupil ghost from the data.  The standard process is to fit the sky background with a 
high-order polynomial across several of the individual CCDs of a DECam exposure, but 
with our narrow-band images this had the effect of removing a portion of the diffuse 
emission as well, leaving negative backgrounds in some subtracted data.  As a result, for our analysis we requested special processing of
the data with options set to fit the sky background with a 2nd-order polynomial across
all detectors in addition to removing the pupil ghost.

\subsection{Data Quality Assessment}\label{sec:data_quality}

To assess the quality of the data listed in Table~2, we examined the 
distribution of the image quality (PSF) and the photometric depth of the
observations, as given by the FITS header keywords SEEING and MAGZERO, 
respectively. These values are determined by the DCP as part of the standard
processing.  
Because the narrow-band DECam surveys of the LMC were performed to surpass 
the existing MCELS data in terms of angular resolution, the DCP 
values for SEEING and MAGZERO are the fundamental measures needed to 
determine if that goal is met. We present histograms of the seeing and 
photometric depth measurements of the observations listed in Table~2 in 
Figures~\ref{seeing_all} \& \ref{magz_all}, and summarize the results in Table~3.

\begin{deluxetable}{lrcc}
\tablecaption{DECam Data Quality}
\tablehead{
 \colhead{Exposure} &
 \colhead{No. Exposures} & 
 \colhead{Median Seeing} & 
 \colhead{Median Photometric Depth}  
\\
 \colhead{Type} & 
 \colhead{} & 
 \colhead{(\arcsec)} &
 \colhead{(mag)} 
}
\tabletypesize{\scriptsize}
\tablewidth{0pt}
\startdata
All Filters        & 1075 & 1.13 &  29.05 \\
\ha\ All           &  165 & 1.37 &  29.85 \\
{[S\,II]} All      &  450 & 1.12 &  29.15 \\
r$^{\prime}$ All   &  460 & 1.13 &  29.00 \\
\hline
All Filters Short  &  458 & 1.14 &  27.10 \\ 
\ha\ Short         &   70 & 1.39 &  26.69 \\
{[S\,II]} Short    &  207 & 1.14 &  27.01 \\
r$^{\prime}$ Short &  208 & 1.13 &  27.56 \\
\hline
All Filters Long   &  590 & 1.13 &  29.70 \\
\ha\ Long          &   95 & 1.26 &  30.30 \\
{[S\,II]} Long     &  243 & 1.11 &  29.70 \\
r$^{\prime}$ Long  &  252 & 1.13 &  29.54 \\
%
\enddata
\label{tab:quality}
\end{deluxetable}

\subsubsection{Seeing}

\begin{figure} [hbt!]
\begin{center}
\includegraphics[scale=0.60]{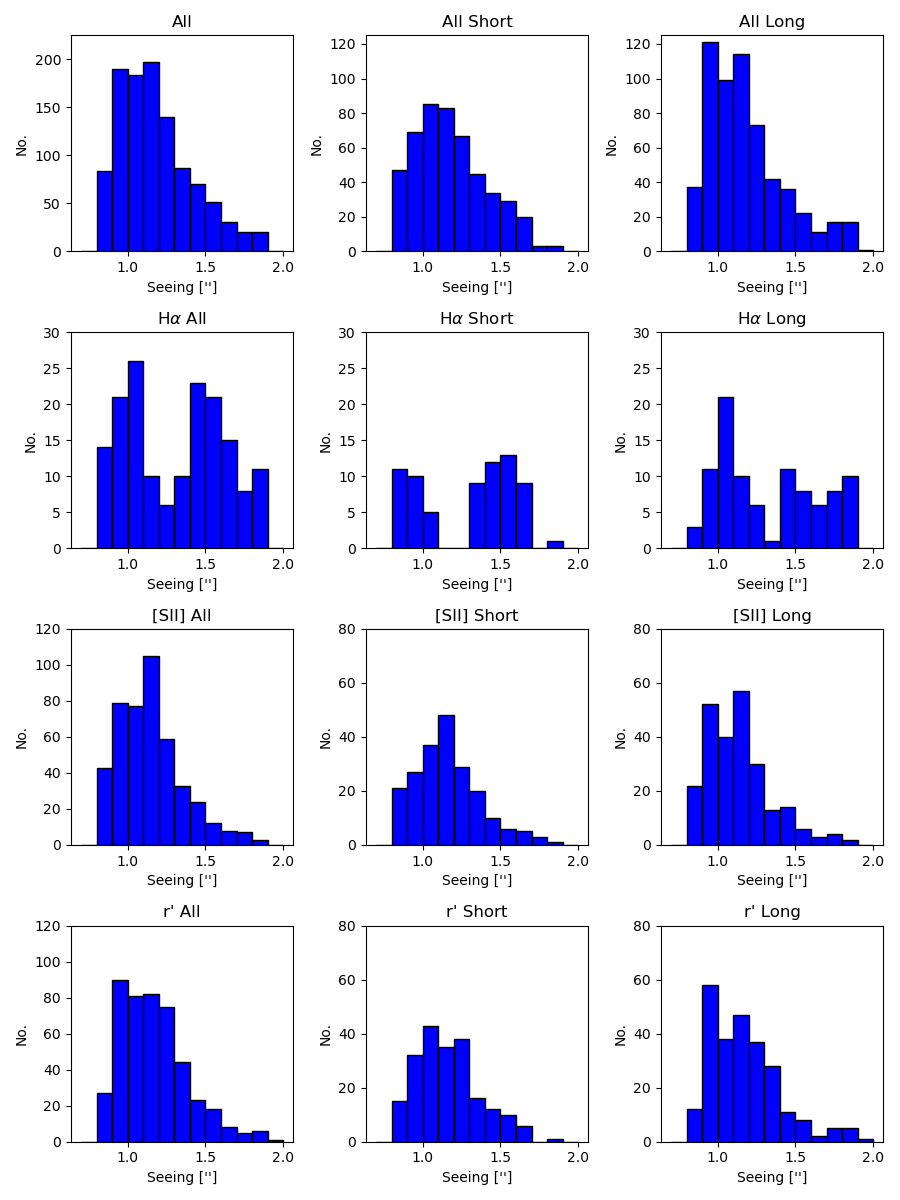}

\caption{\label{seeing_all} Histogram of the seeing value for 
the data as measured by the DCP.  The top row plots the measured seeing for all 
filters and all exposure times ({\bf left}), all filters for short exposure 
times ({\bf center}), and all filters for long exposure times ({\bf right}). 
The second, third, and fourth rows plot the seeing for all exposure times
({\bf left}), short exposure times ({\bf center}), and long exposure times 
({\bf right}), for the \ha, \sii, and r$^{\prime}$ filters, respectively.
The histograms have been divided into 0\farcs1 bins, with the first bin
centered at 0\farcs75 and the last bin centered on 1\farcs95. Note that 
the y-axis for these histograms is not constant.}  
\end{center}
\end{figure}

As seen in the top-left panel of Fig.~\ref{seeing_all}, the seeing
histogram for all exposures and all filters has the most counts in the bin
centered on 1\farcs15, but amplitude is similar to the bins centered on 
0\farcs95 and 1\farcs05.  The overall seeing distribution has a median 
value of 1\farcs13.  Furthermore, the distribution is not Gaussian in shape; 
it has a sharp lower cutoff at 0\farcs8 and a tail that extends to 2\farcs0.  
The seeing distributions for the short and long exposures in all filters 
(top-center and top-right panels, respectively) follow the same general trend. 
We note that the histogram of the short exposures peaks in the bin centered 
on 1\farcs05 and that the long exposures have relatively more instances of
measured seeing $>$ 1\farcs7 than the short exposures.  

In order to investigate the delivered image quality as revealed by seeing
histograms in more detail, we also examine the seeing on a per filter basis 
as shown in Fig.~\ref{seeing_all}.  In general, the seeing distribution of the
\sii\ and r$^{\prime}$ observations follows the distribution seen for all
filters. This is not surprising because, as shown in Table~3, the vast majority
of individual exposures ($\sim85\%$) were taken with the \sii\ and r$^{\prime}$ 
filters contemporaneously in the 2021B semester.  The \ha\ observations, 
taken during the 2018A (LMC\_30Dor field) and 2018B (all other LMC fields),
however, have a distinct bimodal seeing distribution (Fig.~\ref{seeing_all},
second row).  Given that this bi-modality is apparent in both the short and the 
long \ha\ observations, it seems that the seeing was generally poor and
unstable during the nights when the \ha\ data were obtained.  The short
\ha\ observations have a median seeing 0\farcs25 greater than the median
seeing for all short exposures and the long \ha\ exposures have a median seeing 
0\farcs13 greater than the median for all long exposures (see Table~3).

\subsubsection{Photometric Depth}

\begin{figure} [hbt!]
\begin{center}
\includegraphics[scale=0.60]{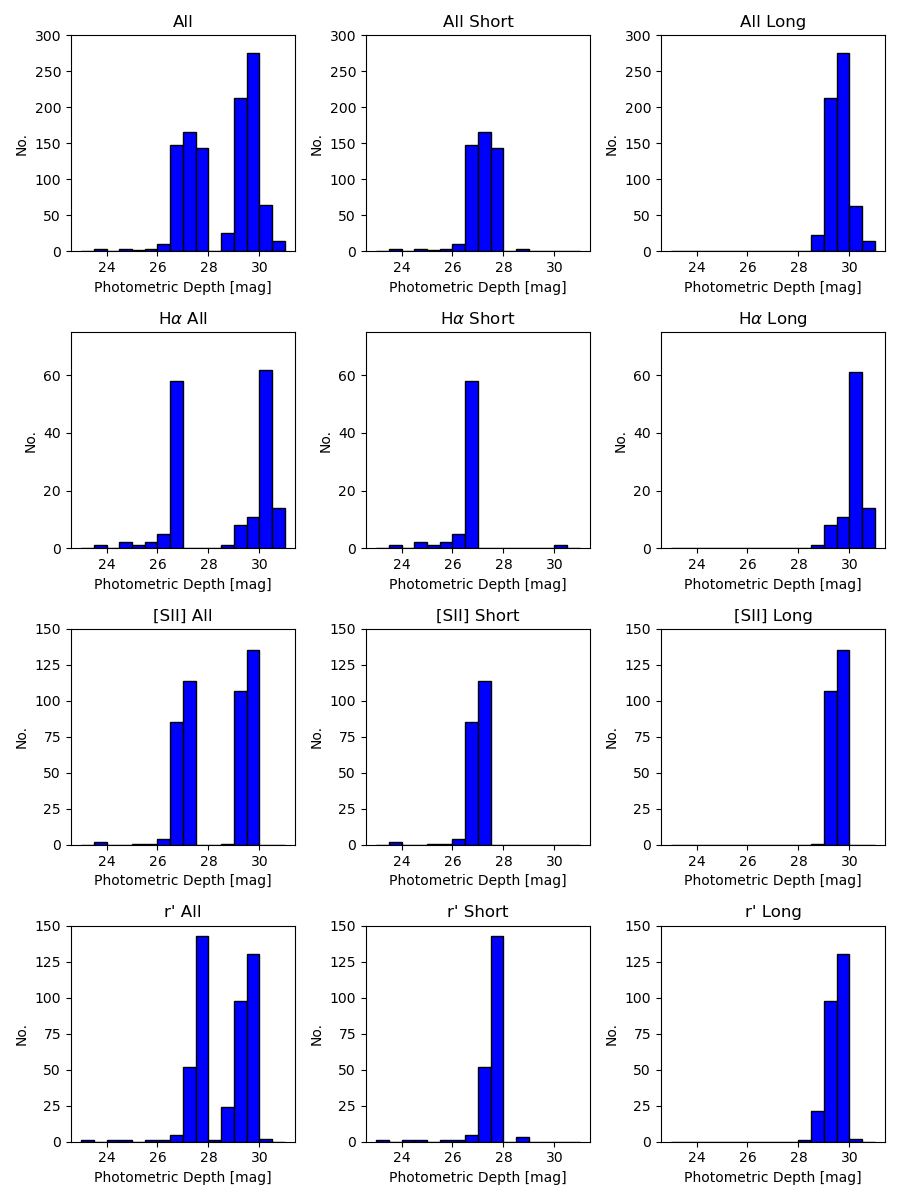}

\caption{\label{magz_all} Histogram of the photometric depth measured by the DCP.  The top row plots the depth for all filters and all exposure times ({\bf left}), all filters for short exposure times ({\bf center}), and all filters for long exposure times ({\bf right}).  The second, third, and fourth rows plot the photometric depth for all exposure times ({\bf left}), short exposure times ({\bf center}), and long exposure times ({\bf right}), for the \ha, \sii, and r$^{\prime}$ filters, respectively. The histograms have been divided into 0.5~mag bins with the first bin centered on 23.25~mag and the last bin centered on 30.75~mag.  Note that the y-limit for these histograms is not constant.}  
\end{center}
\end{figure}

Similar to our analysis of the delivered image quality of the DECam data using 
the seeing histograms, we also investigate the photometric depth of our 
observations.  As seen in the left-hand side panels in Fig.~\ref{magz_all},
the distribution of the photometric depth for all exposure times is bimodal and
independent of the filter used.  This is expected as the short exposures are
not as deep as the long exposures.  When we separate the data into short and 
long exposures and plot histograms of the photometric depth, we find that the 
short and long observations have a median depth of 27.10~mag and 29.70~mag, 
respectively.  For the short exposures, we go deepest with the r$^{\prime}$
observations.  For the long exposures, we are deepest with the \ha\ images, as
measured by the DCP.

We also see from Fig.~\ref{magz_all} that there is more spread in the photometric 
depth values for the short exposures in general and for the \ha\ data in 
particular. Again, this is indicative of the more unstable observing conditions for the observing run when the \ha\ data were obtained.

\subsection{Final Data Set}\label{sec:final_data}

As discussed below in \S\ref{sec:kred}, continuum-subtraction is necessary to 
detect the faintest emission from the ISM, but can be difficult in 
practice.  In order to minimize the effects of poor seeing and poor weather 
(i.e., clouds) on the data presented here, we use the plots shown above in 
Figs~\ref{seeing_all} \& \ref{magz_all} to flag low-quality data and remove it
from our reduction pipeline.  In general, observations with a SEEING value
$>$ 1\farcs45 and observations with MAGZERO values inconsistent with 
the median values determined for the short and long exposures in each filter 
were not processed further{\footnote{We broke this rule only to 
include long \ha\ images with SEEING values of 1\farcs51 and 1\farcs62, 
for fields LMC\_c36 and LMC\_c45 respectively, in our final data set.}}.

\begin{figure} [hbt!]
\begin{center}
\includegraphics[scale=0.60]{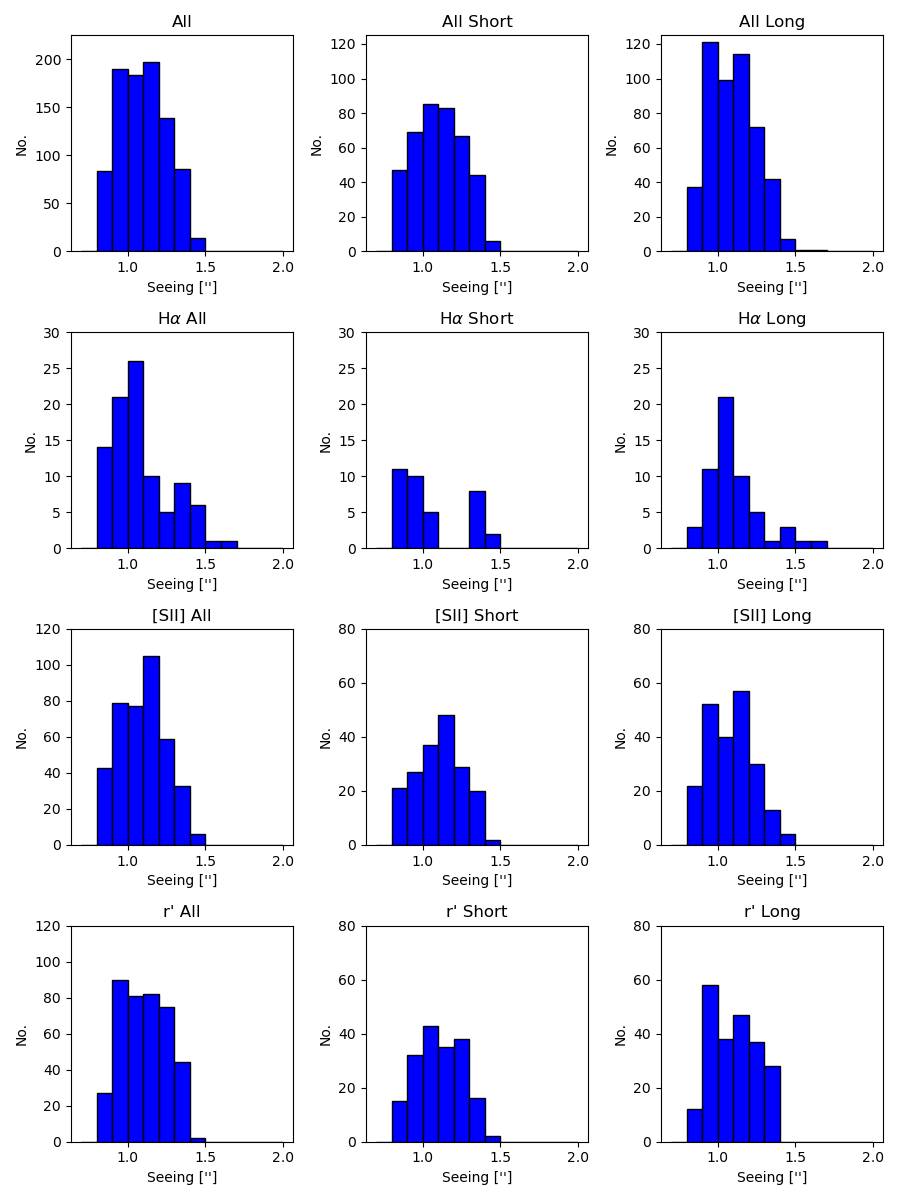}
\caption{\label{seeing_all_cut} Same as Fig.~\ref{seeing_all} except data
with a seeing value $>$ 1\farcs45 have been removed.}  
\end{center}
\end{figure}

\begin{figure} [hbt!]
\begin{center}
\includegraphics[scale=0.60]{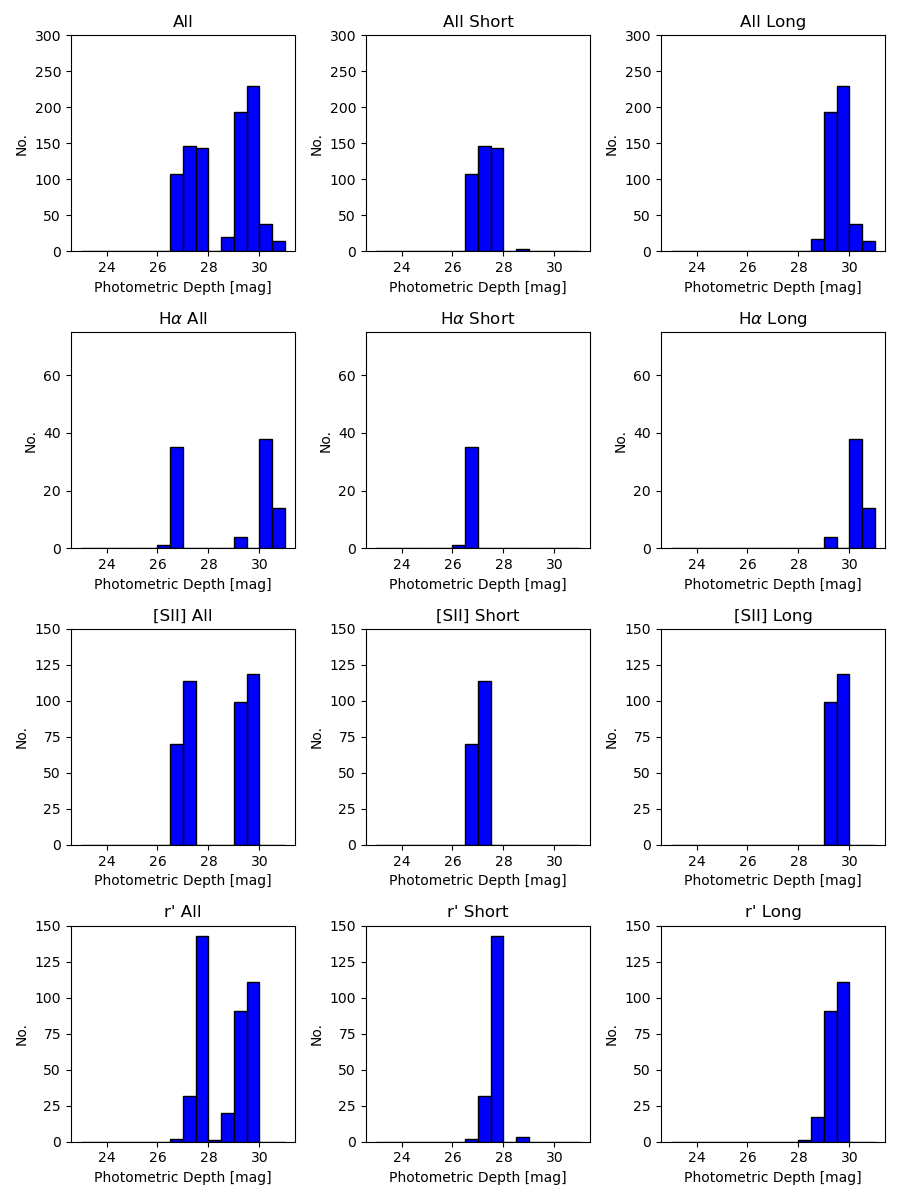}
\caption{\label{magz_all_cut} Same as Fig~\ref{magz_all} except data with
seeing value $>$ 1\farcs45 have been removed.}
\end{center}
\end{figure}

After removing the data with a seeing value $>$ 1\farcs45, we re-examined 
the distribution of the seeing and photometric depth and present those 
revised results in Fig.~\ref{seeing_all_cut} and Fig.~\ref{magz_all_cut} and
summarize them in Table~4.  A comparison between Table~3 and Table~4 shows that,
in general, removal of data based on the seeing value did not drastically improve
the median seeing and median photometric depth of the \sii\ and r$^{\prime}$
observations.  We do, however, see a significant improvement of the measured median seeing
in the \ha\ images.  Before removal of poor seeing data, the median
seeing of the \ha\ observations was 1\farcs37 and after removal the median
seeing value is 1\farcs04.  Likewise, the median photometric depth of all of 
the \ha\ images, after removal of low-quality data, improves by 0.45 mag. 

As seen in Figure~\ref{magz_all_cut}, the histograms of the photometric depth of
the long \ha\ and short r$^{\prime}$ observations have counts in bins centered at
29.25~mag and 28.75~mag, respectively.  The data in these bins come from the 
LMC\_30Dor field (see Table~2) where the long \ha\ and short r$^{\prime}$ 
observations had an individual exposure times of 300~s and 20~s.  Therefore,
in comparison to our more typical long \ha\ and short r$^{\prime}$ of 800~s and
8~s, one expects these fields to have a lower than average depth in the long \ha\ 
observations and greater than average depth in the short r$^{\prime}$ 
observations. 

Finally, for all observations, we did not use any data from the ``S7'' detector 
in DECam.  As mentioned on the DECam web page 
(\href{https://noirlab.edu/science/programs/ctio/instruments/Dark-Energy-Camera/Status-DECam-CCDs}
{https://noirlab.edu/science/programs/ctio/instruments/Dark-Energy-Camera/Status-DECam-CCDs}), 
amplifier B on this detector has an 
unstable gain.  As a result, the background across this detector has a 
discontinuity that affects image combination when matching the sky background
across multiple detectors for image combination.  
Because we are performing sky- and continuum-subtraction to investigate faint
and large-scale nebular emission, we removed this detector from all data
processed by our custom pipeline.

In Table~5, we present the data for the long exposures we process with our custom software, 
described in \S\ref{sec:kred}.  As seen in the table, we did not end up achieving complete 
and uniform coverage of the LMC with the current data.

\begin{deluxetable}{lrcc}
\tablecaption{Revised Data Quality}
\tablehead{
 \colhead{Exposure} &
 \colhead{No. Exposures} &
 \colhead{Median Seeing} & 
 \colhead{Median Photometric Depth}  
\\
 \colhead{Type} &
 \colhead{} & 
 \colhead{(\arcsec)} & 
 \colhead{(mag)}  
}
\tabletypesize{\scriptsize}
\tablewidth{0pt}
\startdata
All Filters        & 894 & 1.09 &  29.07 \\
\ha\ All           &  92 & 1.04 &  30.30 \\
{[S\,II]} All      & 402 & 1.10 &  29.16 \\
r$^{\prime}$ All   & 400 & 1.10 &  29.01 \\
\hline
All Filters Short  & 400 & 1.09 &  27.13 \\ 
\ha\ Short         &  36 & 0.96 &  26.79 \\
{[S\,II]} Short    & 184 & 1.12 &  27.03 \\
r$^{\prime}$ Short & 180 & 1.10 &  27.57 \\
\hline
All Filters Long   & 494 & 1.09 &  29.71 \\
\ha\ Long          &  56 & 1.05 &  30.45 \\
{[S\,II]} Long     & 218 & 1.07 &  29.74 \\
r$^{\prime}$ Long  & 220 & 1.10 &  29.56 \\
\enddata
\label{tab:new_quality}
\end{deluxetable}

\begin{deluxetable}{lcccccl}
\tablecaption{Revised Observation Summary}
\tablehead{
 \colhead{Field ID} & 
 \colhead{R.A. (J2000)} & 
 \colhead{Decl. (J2000)} & 
 \multicolumn{3}{c}{Filters} &
 \colhead{Complete\tablenotemark{a}}
\\
 \colhead{} & 
 \colhead{} & 
 \colhead{} & 
 \colhead{DES r$^\prime$} &
 \colhead{N662}  &
 \colhead{N673}  &
 \colhead{}
\\
 \colhead{} & 
 \colhead{(deg)} & 
 \colhead{(deg)} & 
 \colhead{(s)} & 
 \colhead{(s)} & 
 \colhead{(s)} &
 \colhead{}  
}
\tabletypesize{\scriptsize}
\tablewidth{0pt}
\startdata
LMC\_30Dor & 84.640830 & $-$69.085860 &  \nodata         &  4 $\times$ 300.0 & \nodata           & N (S,r)   \\
LMC\_c01   & 73.591856 & $-$66.807029 & 11 $\times$ 60.0 &  \nodata          & 12 $\times$ 800.0 & N (H)   \\
LMC\_c02   & 75.395313 & $-$65.915355 & 10 $\times$ 60.0 &  \nodata          & 10 $\times$ 800.0 & N (H)   \\
LMC\_c03   & 81.041359 & $-$65.749791 & 12 $\times$ 60.0 &  \nodata          & 12 $\times$ 800.0 & N (H)   \\
LMC\_c30   & 72.804846 & $-$68.628810 &  6 $\times$ 60.0 &  2 $\times$ 800.0 &  7 $\times$ 800.0 & N (H,S,r) \\
LMC\_c32   & 73.662449 & $-$70.431310 &  4 $\times$ 60.0 &  2 $\times$ 800.0 &  6 $\times$ 800.0 & N (H,S,r) \\
LMC\_c35   & 75.925342 & $-$67.750762 & 12 $\times$ 60.0 &  4 $\times$ 800.0 & 12 $\times$ 800.0 & Y         \\
LMC\_c36   & 76.827761 & $-$69.387311 & 24 $\times$ 30.0 &  2 $\times$ 800.0 & 24 $\times$ 400.0 & N (H)   \\
LMC\_c37   & 77.475267 & $-$66.306975 &  2 $\times$ 60.0 &  \nodata          &  2 $\times$ 800.0 & N (H,S,r) \\
LMC\_c38   & 77.761190 & $-$71.196145 &  8 $\times$ 60.0 &  5 $\times$ 800.0 &  6 $\times$ 800.0 & N (S)   \\
LMC\_c39   & 80.141220 & $-$67.952984 & 12 $\times$ 60.0 &  4 $\times$ 800.0 & 12 $\times$ 800.0 & Y         \\
LMC\_c41   & 80.971450 & $-$69.756543 & 28 $\times$ 30.0 &  5 $\times$ 800.0 & 24 $\times$ 400.0 & Y         \\
LMC\_c42   & 81.831476 & $-$66.471265 & 10 $\times$ 60.0 &  5 $\times$ 800.0 & 10 $\times$ 800.0 & Y         \\
LMC\_c43   & 81.923156 & $-$71.556051 &  7 $\times$ 60.0 &  5 $\times$ 800.0 & 10 $\times$ 800.0 & Y         \\
LMC\_c45   & 83.928142 & $-$68.444092 & 24 $\times$ 30.0 &  2 $\times$ 800.0 & 24 $\times$ 400.0 & N (H)   \\
LMC\_c47   & 85.215270 & $-$70.080852 & 28 $\times$ 30.0 &  5 $\times$ 800.0 & 25 $\times$ 400.0 & Y         \\
LMC\_c48   & 85.481748 & $-$66.961984 & 11 $\times$ 60.0 &  \nodata          & 11 $\times$ 800.0 & N (H)   \\
LMC\_c49   & 86.134066 & $-$71.855063 & \nodata          &  6 $\times$ 800.0 &  \nodata          & N (S,r)   \\
LMC\_c50   & 88.158561 & $-$68.600023 & 11 $\times$ 60.0 &  \nodata          & 11 $\times$ 800.0 & N (H)   \\
LMC\_c51   & 89.083554 & $-$70.376535 & \nodata          &  5 $\times$ 800.0 &  \nodata          & N (S,r)   \\
\enddata
\tablenotetext{a} {Completeness of the observations for the final dataset as shown in Figure~\ref{lmc-overview}. ``Y'' 
indicates that at least 2/3 of the observations for a field were completed and passed the quality assessment. ``N'' 
indicates that fewer than 2/3 of the observations for a field were completed and/or failed the quality assessment. The
incomplete observations for the individual filters are labeled as ``H'' (\ha), ``S'' ([S\,II]), and ``r'' (r$^{\prime}$).}
\label{observations}
\end{deluxetable}

\section{Data Processing}\label{sec:kred}

The survey data were obtained over a number of nights and under varying conditions with an instrument containing 60 CCDs with significant gaps between individual detectors.  Our goal in processing these data was to produce mosaicked versions of these images and to accurately represent diffuse \ha\ and \sii\ emission (on large and small scales) at  brightness levels that are significantly lower than that observed from the night sky. This presents significant challenges since on the one hand (a) all of the images contain significant contribution from stars, which range from isolated objects to clusters, and (especially in the bar of the LMC) to the diffuse light which is brighter than the emission line gas, and (b) the band pass of the $\rm r^\prime$-band filter is sufficiently broad that these images contain contributions from both \ha\ and \sii, albeit at low levels due to the shorter exposure times (see Fig~\ref{filters}).  Our goals and the associated challenges are very different from those who wish to measure transient phenomena in sets of images, or to study starlight.

To address these goals, and after significant experimentation with alternatives, we have developed a processing pipeline that is designed to be optimized for accurately assessing the diffuse emission that pervades the LMC. The pipeline is Python-based, but makes extensive use of the {\sc SWARP} software package \citep{SWARP} to re-project and mosaic images.  The principal steps of this procedure are outlined below:

\begin{itemize}
\item We begin with the data (ooi images) reduced by the DCP that has passed 
the data quality checks discussed in sections \S\ref{sec:data_quality} and 
\S\ref{sec:final_data}.

\item We re-scale all of the data to a common (stellar) magnitude scale where 1 count (DN) corresponds to that expected for a 27th magnitude star. The re-scaling is based on the flux conversion to magnitudes provided by the DCP which for the declinations appropriate to the Magellanic Clouds is based on the Gaia (3rd early release) G band catalog. To zeroth-order, this conversion means that one can simply subtract any two images taken at the same position from one another to produce a continuum-free difference image.   
We also remove a single background from all the CCD images that comprise each exposure, providing a zeroth-order subtraction of the sky background.  

\item We create a grid of 4$\times$4 overlapping tiles for each field, each 0\fdg67  across. This is basically a convenience that allows for producing a set of uniform data products. The remaining analysis is carried on the individual CCD images that are part of each tile.

\item For each tile and for each set of exposures with filter and 
exposure time, we measure the difference in flux levels in the overlap 
regions, and use this to adjust the background levels on each CCD. Specifically, we use {\sc SWARP} to project the individual CCDs onto a common astrometric frame. We then estimate the difference in (sky) background in each of the two overlapping CCD images (from the mode of the difference between the two images in the overlap region).  
Because the typical number of overlaps greatly exceeds the number of images to 
be combined, we fit the overlaps by assigning weights to each 
measurement and use a least-squares procedure to minimize the overlap differences assuming a single 
background be added or subtracted from each exposure, namely

    \begin{equation}
    \Xi = \Sigma_{ij} w_{ij} \left (   \Delta_{ij} - (b_i - b_j) \right )^2
    \end{equation}

where $w_{ij}$  is a weight based on the size of the overlap, $\Delta_{ij}$ is the calculated difference in the fluxes, and $b_i$ is the extra background to be added or subtracted.

\item We then use {\sc SWARP} to create tile images on the same world coordinate system for each exposure and filter.  All of the images at this point in the processing contain a combination of continuum (from stars) and line emission (from gas).  The stars are of roughly equal brightness in all of the images, but the relative amount of line emission depends upon the filter bandpass.

\item  To create ``pure'' emission-line images, we first create ``emission-line free continuum images'' from the $\rm r^\prime$-band images and subtract
these images from the \ha\ and \sii\ images.

\end{itemize}

The basic process is straightforward.  The $\rm r^\prime$-band image contains continuum emission from stars as well as line emission from  \ha\ (and \nii) and from \sii, e.g.,

\begin{equation}
    r^\prime=r^\prime_{cont}+r^\prime_{H\alpha}+r^\prime_{[S~II]}
\end{equation}

\noindent
All  of the images have been scaled so that stars are equally bright in the images.  Consequently, the counts (DN) from \ha\ emission in the r$^\prime$ band images is only a fraction of  the \ha\ counts (DN) in the  \ha\ images, that is

\begin{equation}
    r^\prime_{H\alpha}=\frac{\Delta\lambda_{H\alpha}}{\Delta\lambda_{r^\prime}} H\alpha
\end{equation}

where $\Delta\lambda$ is the effective band pass, approximately the FWHM of each filter.

\noindent

Similarly for \sii,

\begin{equation}
   r^\prime_{[SII]}=\frac{\Delta\lambda_{[S~II]}}{\Delta\lambda_{r^\prime}} [SII]
\end{equation}

So one can now create a line free $\rm r^\prime$-band image by subtraction, e.g.,

\begin{equation}
    r^\prime_{cont}\propto \left ( r^\prime-\frac{\Delta\lambda_{H\alpha} H\alpha  + \Delta\lambda_{[S~II]} [SII]}{\Delta\lambda_{r^\prime}} 
    \right )
\end{equation}

However, in the process of subtracting the line emission from the $\rm r^\prime$-band image, we have also removed some of the continuum emission.  In order to produce a final emission-line subtracted image with
stars of the same apparent brightness as in the original image, one must renormalize:

\begin{equation}
    r^{\prime}_{cont}=  \frac{\Delta\lambda_{r^\prime} r^{\prime} -(\Delta\lambda_{H\alpha} H\alpha  + \Delta\lambda_{[S~II]} [SII])}{\Delta\lambda_{r^\prime}-(\Delta\lambda_{H\alpha} + \Delta\lambda_{[S~II]})} 
\end{equation}

Thus, given an $r_{cont}$ image, one can in principle produce a pure \ha\ or \sii\ image by simple subtraction. This process is not perfect for a variety of reasons, which include that (a) there are color corrections associated with the differences in central wavelengths, and these can vary from field to field, and (b) the $\rm r^\prime$-band images were typically taken at different times from the narrow-band images and so the stellar PSFs differ between the narrow and $\rm r^\prime$-band images.  Nevertheless, this process produces superior continuum images for using to subtract from the emission-line images.  

In principle, further improvements could be made, especially by convolving the images to a common PSF prior to image subtraction, but the current version of the processing pipeline does not include this.\footnote{The current version of the processing pipeline, which we refer to as {\sc kred}, can be found on github at \href{https://github.com/kslong/kred}{https://github.com/kslong/kred}.  We would be pleased to have others make use of it, and/or to help improve it further.}  As described in \S\ref{sec:final_data}, we attempt to at least partially ameliorate this problem by limiting the data we use to have a seeing value $<$ 1\farcs45.

As noted earlier, as part of the processing the images, we rescaled them all based on G-band magnitudes in the Gaia catalog. Given this scaling, the narrow-band flux in the final \ha\ and \sii\ images should be given by

\begin{equation}
F_{\lambda_o} = DN  \frac{1}{A(\lambda_o)\lambda_o} \int {A(\lambda)}{\lambda}  F_{\lambda, 27} d\lambda 
\end{equation}

\noindent
where $A$ represents the effective area of the DECam system, allowing for the transmission of the atmosphere, the reflectivity of the mirror system of the telescope and any filters used in the observation, and $F_{\lambda, 27}$ is the flux per unit wavelength above the atmosphere for a 27th magnitude star. For the narrow-band \ha\ and \sii, and for a wavelengths in region where the transmission through the narrow-band filter is near the peak, the flux conversion reduces to

\begin{equation}
F_{\lambda_o} = DN  \ F_{\lambda, o, 27}\Delta \lambda
\end{equation}

\noindent
where $\Delta \lambda$ is the filter bandpass, 160 and 100 \AA\ in the cases of the \ha\ and \sii\ filters, respectively.  The 3rd data release of Gaia data includes spectra of a large number of stars that appear in the images, and based on the spectra, estimates of the effective temperatures and gravities of many of these stars.  To improve on the flux calibration  provided by DCP, we have selected stars with temperatures near 10,000 K  and g-band magnitudes of order 16 and matched them to objects for which we have measured net counts using aperture photometry.  Scaled to 27th magnitude, we find that the emission line flux corresponding 1 DN in the \ha\ and \sii\  images is \EXPU {2.4}{-18 }{\FLUX} and \EXPU {2.1}{-18 }{\FLUX}, respectively.   Given the pixel size of 0\farcs27, these fluxes correspond to surface brightnesses of \EXPU{3.3}{-17}{\FLUX~arcsec^{-2}} and \EXPU{2.9}{-17}{\FLUX~arcsec^{-2}}, respectively.  The \ha\ surface brightness limit corresponds to an emission measure of 16.5~cm$^{-6}$~pc.  Extended objects with excess of 1 DN are readily visible in the final images.

\section{A Sampling of Initial Results}\label{sec:results}

We consider the currently available processing of the data to be preliminary, but the superior resolution and overall depth and quality of the DECam survey data are already apparent.  In this section, we provide some examples of the new data for various types of nebulae and compare to the previous MCELS survey data.  Of course, these examples barely scratch the surface; the ultimate goal is to provide the full data set to the community, allowing larger, global studies to be performed, taking advantage of the improved spatial resolution.

\begin{figure} [hbt!]
\begin{center}
\includegraphics[scale=0.50]{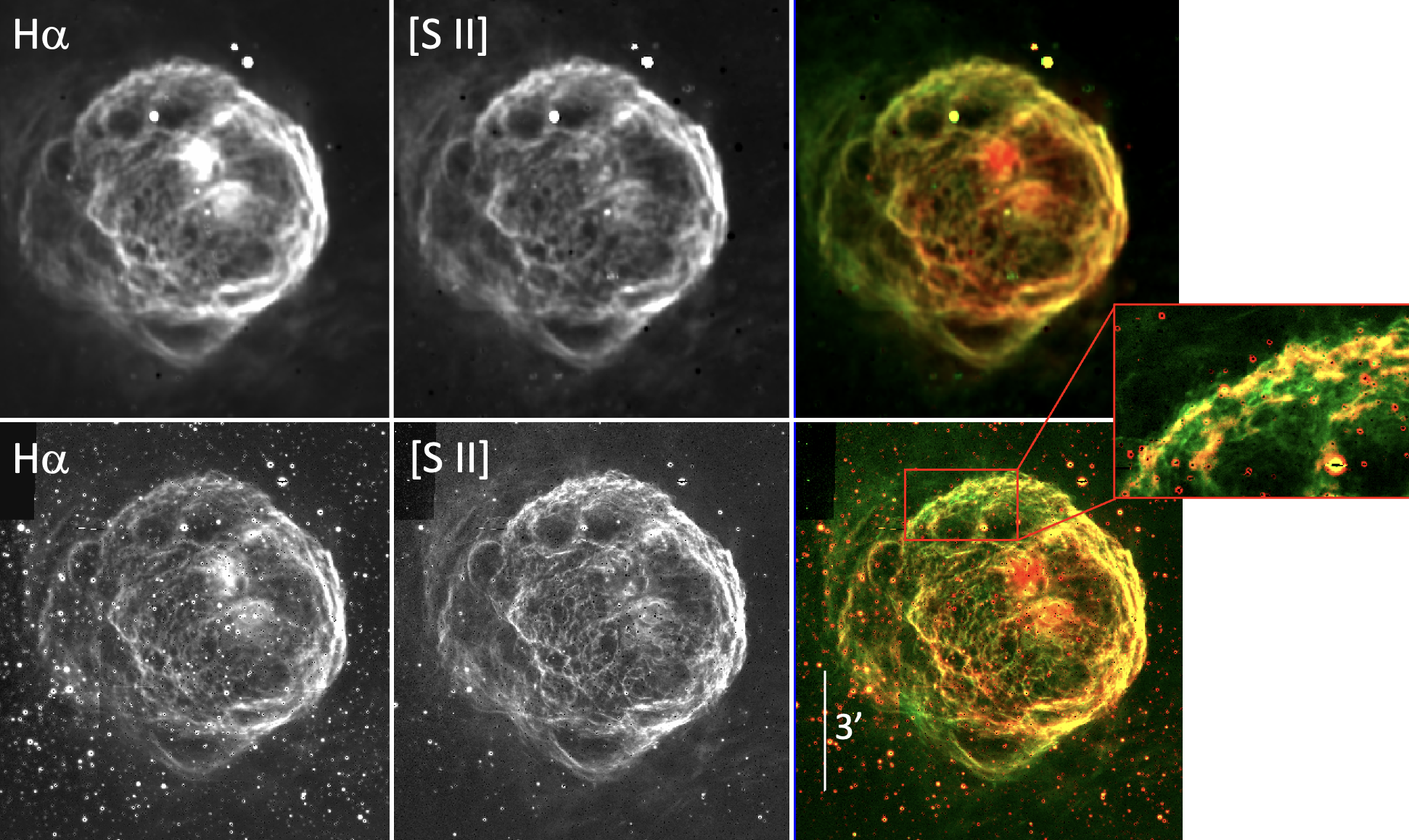}
\caption{\label{N70} A comparison of MCELS and DeMCELS data for the N70, a superbubble $\sim$7\farcm8 in diameter, and located in an isolated region in the eastern LMC. Top row shows MCELS H$\alpha$ and \sii\ images in black and white (linear stretch), with a color combination at right (red: H$\alpha$; green: \sii, log scaling).  Yellow indicates both ions are strong.  The bottom row shows the same sequence, but for our DECam data. The scale is shown in the lower right color panel. The inset at right shows detail of the northern rim.}
\end{center}
\end{figure}

\subsection{Bubbles and Superbubbles}

Fig.~\ref{N70} shows the region surrounding N70 \citep{Henize56}, an isolated nebular superbubble in the eastern LMC, also known as DEM L301 \citep{Davies76}. N70 has a diameter of 7\farcm8 ($\sim$115 pc) and surrounds the stellar association LH114 \citep{Lucke70}.  
N70 is an X-ray emitter \citep{Chu90, wang91a, Zhang2014} but is relatively weak.  

Optical spectroscopic observations show the nebula to be expanding at $\sim40 ~ \kms$ \citep[][and references therein]{Chu88a}.  Optical spectra also show \siiL\ to \ha\ ratios to be variable but averaging $\sim$0.3, somewhat enhanced above the value expected for bright photoionized nebulae.  Shock heating from strong stellar winds and/or previous supernovae within the shell are possible sources, but early modeling by \citet{Dopita81} did not conclude shock heating was required. 
It was surprising that far ultraviolet observations of stars within the superbubble showed an excess of \oviL\ absorption compared with non-superbubble sight lines \citep{Danforth06}.  Since shock velocities in excess of 150~$\kms$ are needed to produce \ovi, a model involving thermal conduction from nebular interfaces with the hot interior gas seems favored.

The stellar content exciting the superbubble emission has been analyzed extensively and compared with other superbubbles and comparable \hii\ regions \citep{Oey96a, Oey96b}.  The photoionizing input for N70 is apparently dominated by a few very early type O stars \citep{Oey96b} and calculations indicate there should be more than sufficient ionizing radiation to photoionize the entire nebula \citep{Oey96c, Skelton99}.  
In fact, an embarrassingly little of the available energy is needed, leading to the suggestion that much of the ionizing radiation must escape the nebula. The differing optical nebular morphology between classical \hii\ regions and superbubbles like N70 does not appear to be driven by the stellar content, as color magnitude diagrams (CMDs) show very similar behavior \citep{Oey96b}.  
Fitting the CMD for N70 with a Salpeter IMF implies that one or more very massive association members should have already exploded in the region, raising the likelihood that a hybrid model involving both photoionization and shocks play a role in exciting the nebula, a concept that was well-studied and modeled by \citet{Oey00}.

\citet{Skelton99} used the Rutgers Fabry-Perot on the CTIO 1.5m to obtain CCD imagery of N70 with very narrow effective filter bandpasses to sample a wider range of emission lines, including \oiii\ $\lambda$ 5007, and cleanly separating \ha\ from \nii\ $\lambda\lambda$ 6548,6583, but at a seeing-limited 2\farcs2 resolution.\footnote{While the \citet{Skelton99} data separated \ha\ and \nii, the spatial distribution of these two lines is quite similar.}   
Those observations show the nebular ionization structure (including \oiii) to good advantage,  
and indicate a mix of radiative and collisional (shock) processes is likely responsible for ionizing and exciting the nebular shell, with the fraction of each changing outward from the center (more photoionization) to the outer rim (more evidence of shocks).  This highlights the advantage of eventually adding \oiii\ to our own survey as an additional nebular ionization diagnostic.

The comparison in Fig.~\ref{N70} shows MCELS data in the top three panels and our new DECam survey images below.  DECam resolves the filamentary structure of N70 at higher angular resolution, showing the nebular structure to be even more filamentary than seen previously.   
This may help explain why so little of the available ionizing flux is captured in the observed nebula.  The brightening of the nebula on the west side likely indicates the expansion is encountering denser material on that side. 
The color variations in the lower right panel show an even finer separation of regions with enhanced \sii\ compared with \ha, highlighting more dramatically those filaments dominated by shock heating, including a cluster of filaments just east of north on the north rim  (see inset), faint filaments to the southeast, and an apparent and impending blow-out from the shell in the south.  
Targeting specific regions for further spectroscopic studies will allow the full extent of shock heating versus photoionization to be investigated, providing more detailed constraints to modeling such nebular systems.  Of course, the power of the new survey is that such studies can be performed on numerous nebulae, sampling a wide range of parameter space.

\begin{figure} [hbt!]
\begin{center}
\includegraphics[scale=0.60]{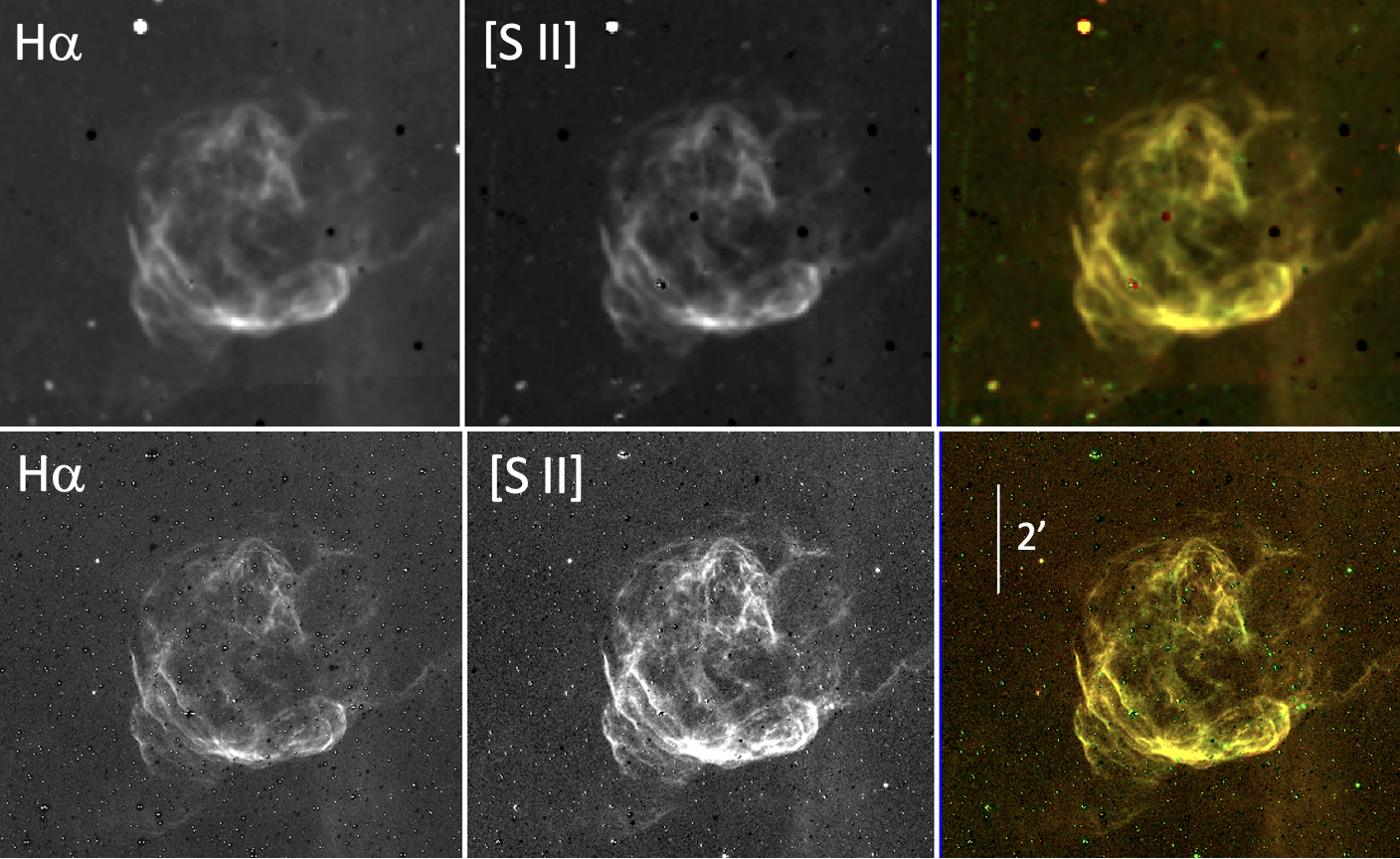}
\caption{\label{DEML204} A comparison of MCELS and DeMCELS data for the faint SNR 0527-6549, aka DEM L204.  Presentation is the same as in Fig.~\ref{N70}.  This is a low surface brightness SNR in field c42 (tile 12). Note the subtle color (ratio) variations in the DECam image at lower right.}
\end{center}
\end{figure}

\subsection{Supernova Remnants}

One of our team's primary goals is to improve upon observations of LMC SNRs, 
particularly the population of larger, fainter SNRs that are associated 
with later stages of SNR evolution leading to merging with the ISM.  This survey allows us to use \sii/\ha\ ratios to verify candidate SNRs suggested in various wavelength regimes, and to uncover new SNRs.

A handful of the brightest LMC SNRs have been observed individually in depth and at many wavelengths, including some of the most famous SNRs observed with HST.  However, many of the fainter or lesser known objects have received little attention and have relatively poor quality imagery available.  The broad spatial coverage of DeMCELS addresses this set of SNRs by providing improved imagery in \ha\ and \sii.

Fig.~\ref{DEML204} shows one such object, DEM L204, where previous MCELS and DeMCELS data are compared.  This faint SNR lies in an isolated region in the northern LMC. The DECam images resolve the nebular structure in significant detail, showing a partial shell of emission brightest in the south and open to the west, where faint filaments extend beyond the main shell. At high resolution, variations in ratio for individual filaments are apparent as color variations in the panel at lower right.

\begin{figure} [hbt!]
\begin{center}
\includegraphics[scale=0.65]{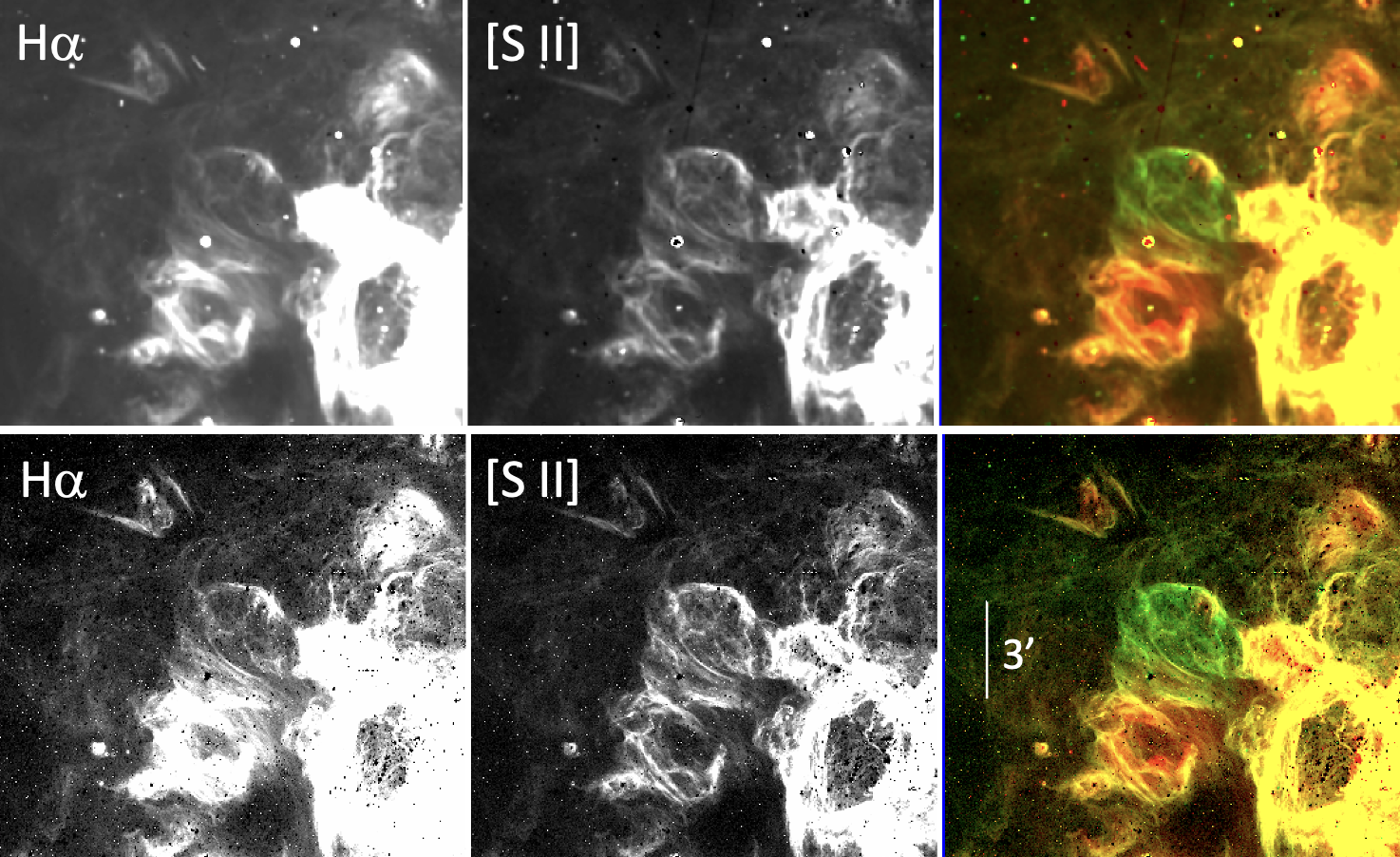}
\caption{\label{SNR0523} A comparison of MCELS and DeMCELS data for the faint SNR 0523-6753.  Presentation is the same as in Fig.~\ref{N70}.  The SNR stands out by way of it's relatively brighter \sii\ emission despite the complexity of the emission in this region. }
\end{center}
\end{figure}

Another type of SNR where DeMCELS data are superior are in regions of complex emission.
SNR 0523-6753 
is a little-studied SNR on the NE edge of the complex N44 emission region \citep{Chu1993}.
Fig.~\ref{SNR0523} shows the comparison of MCELS and DeMCELS data for this region.  Despite a fair amount of adjacent complexity in emission, the \sii\ image shows details of the filamentary structure of this object clearly.

Even in very complex regions of emission, DeMCELS makes it possible to discern embedded SNR emission.
The Honeycomb SNR (aka SNR 0535-6918) is a region of \sii\ bright loops first identified by \citet{Wang92} 
in a complex field not far from SN 1987A (also contained in this tile). This region contains ten or more loops of emission with sizes of  $\sim$12\arcsec, corresponding to 2.8 pc at the distance of the LMC.  \citet{Chu95} used bright X-ray emission, non-thermal radio emission, and high \sii\ to \ha\ ratios to identify this structure as resulting from a supernova shock front, perhaps propagating through a porous region of dense interstellar gas. \citet{Meaburn10} modeled the spatial and kinematic structure of this object and suggested two possible scenarios: one in which the supernova occurred within the edge of a giant LMC shell, and one in which the loops are produced by a precessing jet from a binary microquasar, as in SS 433.    

\begin{figure} [hbt!]
\begin{center}
\includegraphics[scale=0.55]{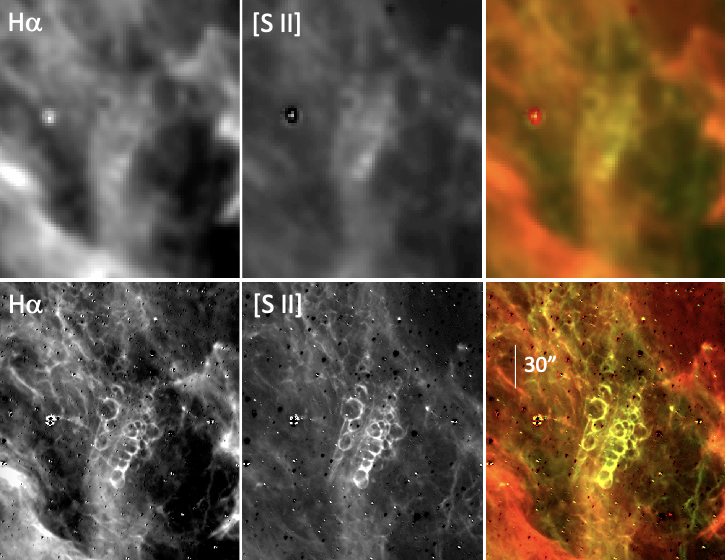}
\caption{\label{Honeycomb} A comparison of MCELS and DeMCELS data for the SNR 0535-6918, aka the Honeycomb, within the outer regions of 30 Dor.  Presentation is the same as in Fig.~\ref{N70}. DECam resolution is particularly effective in separating different kinds of emission in complex regions like this.}
\end{center}
\end{figure}

The large area covered by the DECam pointing in this region (Fig.~\ref{Honeycomb}) allows us to examine both the larger context of the SNR's surroundings, and to examine the loops or ``bubbles'' themselves in detail. If the Honeycomb is part of a larger structure with varying density, as suggested by \citet{Chu95}, it would be useful to examine both the loops themselves and other \sii\ bright features in the vicinity. By using the \sii\ to \ha\ ratios to identify possible features associated with the Honeycomb structure, one could perform follow-up high-resolution velocity studies of these features, building on the work of \citet{Meaburn93}, to see whether, \eg, their line profiles are consistent with a common physical origin.  Such work may help to clarify whether the Honeycomb can be associated with other nearby shocked gas, indicating its status as part of a larger structure. In particular, it would be useful if fainter filaments associated with the parts of the SNR expanding into lower-density surroundings can be identified. SNRs are thought to lose many of their prominent observational characteristics when expanding into rarefied ISM such as the interior of a superbubble. Identifying fainter, more extended structure in this instance may help to quantify the contributions of such otherwise invisible SNRs to the energy and hot gas component in active stellar regions that produce similar low-density conditions. 

\begin{figure} [hbt!]
\begin{center}
\includegraphics[scale=0.25]{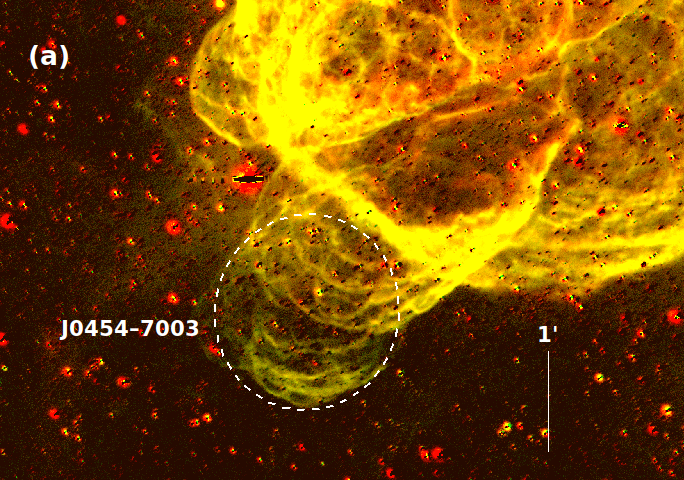}~
\includegraphics[scale=0.25]{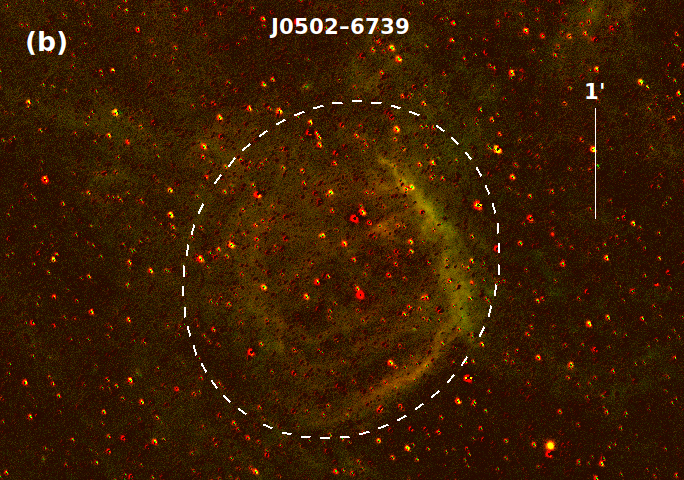}
\includegraphics[scale=0.25]{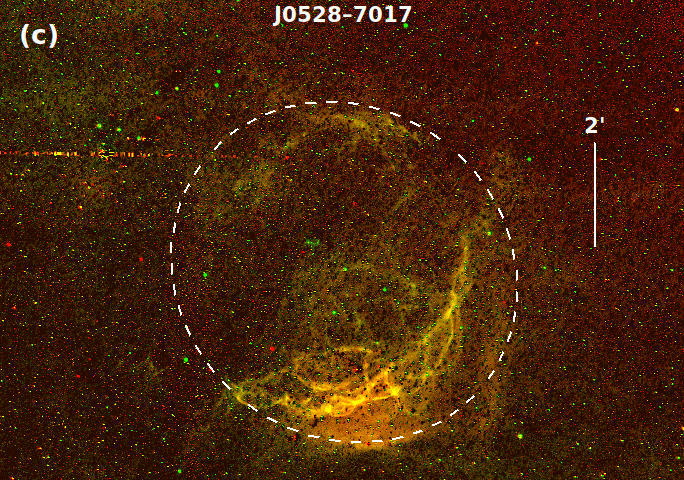}
\caption{\label{YewSNR} DeMCELS data for three SNRs studied by \citet{Yew21}.  Shown here are (a) Yew 3 (SNR 0454–7003), (b) Yew 6 (SNR 0502–6739), and (c) Yew 12 (SNR 0528-7017).   Ellipses indicate the listed sizes of the SNRs from Table 1 of that paper: for Yew 3 the longest axis is 129\arcsec, for Yew 6, 190\arcsec, and for Yew 12, 380\arcsec. \ha\ emission is shown in red and \sii\ in green.}
\end{center}
\end{figure}

Finally, the DECam survey also allows detailed follow-up examination of faint and little-studied SNR candidates, many suggested by observers at radio or X-ray wavelengths.  The MCELS dataset, combined with optical spectroscopy, was used by \citet{Yew21} to study three SNRs and 16 SNR candidates in the LMC.  Fig.~\ref{YewSNR} shows three of these objects in the DeMCELS data.  Yew 3 (SNR 0454–7003) lies at the edge of superbubble DEM L25 (aka N185), which shows some indications of shock activity overall \citep{Oey02}.  Echelle studies of the DEM L25 superbubble by \citet{Zhang2014} show expansion velocities of up to 200 km/s in part of this object, consistent with the presence of SNR shocks, as well as diffuse X-ray emission.  In DeMCELS, the SNR candidate shows up with relatively bright \sii\ emission, as with SNR 0523-6753 mentioned above; but the increased resolution of the DECam survey shows additional structure within the larger superbubble, including other \sii-enhanced regions. Using DeMCELS data to produce a high-resolution \sii/\ha\ ratio map, in combination with existing optical spectra and/or detailed X-ray imaging spectroscopy, could help to more narrowly constrain the contribution of shock structures to the overall superbubble emission. 
Yew 6 (SNR 0502–6739) is an SNR candidate with a comparatively low \sii/\ha\ ratio of 0.55 \citep{Yew21}, which the authors point out may be due to dilution from the overlapping \hii\ region. In the higher resolution DeMCELS data, we can see that the brightest and best defined \sii\ features correspond well with the boundaries estimated from MCELS images, but it also displays  additional faint filamentary structure.  In addition, the well-defined filaments may allow better measurement of shock-heated emission above that of the \hii\ region. Similarly, Yew 12 (SNR 0528-7017) shows fine filaments in the DECam data that were unresolved in MCELS.

\begin{figure} [hbt!]
\begin{center}
\includegraphics[scale=0.45]{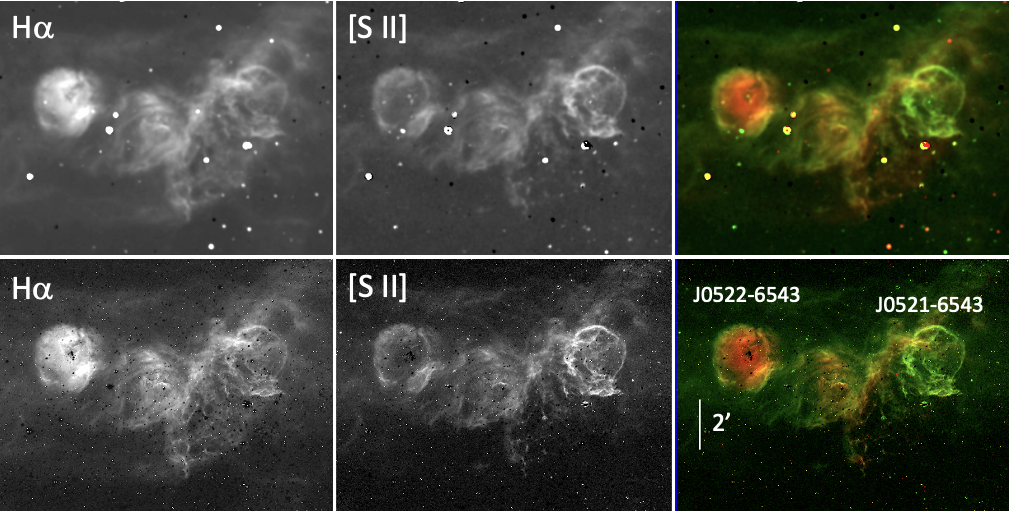}
\caption{\label{SNR0521reg} 
A comparison of MCELS and DeMCELS data for the region near SNR J0521-6543, aka DEM L142, which is the well-defined filamentary shell at right.  Presentation is the same as in Fig.~\ref{N70}. SNR J0521-6543 is $\sim$170\arcsec\ in diameter \citet{Maggi16}. The similar sized shell at left is J0522-6543, an SNR according to \citet{bozzetto23}.  However, this shell is center-filled with \ha\ and does not resolve into crisp filaments at DECam resolution.  The centrally-placed star cluster makes it likely that this nebula is a stellar bubble. See text for details.}
\end{center}
\end{figure}

Some objects suggested to be SNRs by multiwavelength data may not pan out under closer inspection at optical wavelengths.
Fig.~\ref{SNR0521reg} shows a moderately wide view of region of complex emission in the northern LMC.
SNR J0521-6543 (DEM L142) was noted as a possible SNR candidate from MCELS data (e.g., \citealt{Williams2009}) 
and was listed by \citet{Maggi16} as a confirmed SNR. In the DECam images, the \sii\ bright circular shell stands out nicely from the background \ha\ emission.  Not previously noted, however, are the complex loops seen in both emission lines along and below the circular rim of the SNR, reminiscent of the structure in the Honeycomb SNR. This raises the possibility, as suggested for the Honeycomb SNR, that SNR shocks are traveling through a larger structure of porous gas. 

A circular region of about the same size as SNR J0521$-$6543 appears at left in Fig.~\ref{SNR0521reg}, and shows a circular rim of enhanced \sii\ emission.  This feature aligns with a shell-like radio feature noted by \citet{bozzetto23} in ASKAP and ATCA images. Its shell-like radio morphology, and its radio spectral index of $-0.51 \pm 0.05$, typical of SNRs, led the authors to suggest this region as a confirmed SNR J0522-6543.  They also note an enhanced  \sii\ to \ha\ ratio of 0.4 from MCELS images.  However, the DECam images of this region show bright central \ha\ emission with a diffuse surrounding ring of enhanced \sii; as viewed at DECam resolution, the ring does not show the sharp filamentary structure seen in most LMC SNRs (including J0521-6543 at right). This diffuse morphology more strongly suggests that the \sii\ emission marks the outer boundary of photoionized gas in an \hii\ region.  Although continuum has been subtracted in  Fig.~\ref{SNR0521reg}, residuals from a central star cluster can be readily seen.

Adding to the interest of this optical complex is the slightly enhanced \sii\ in an irregular filamentary feature that lies between the SNR and SNR candidate. 

\begin{figure} [hbt!]
\begin{center}
\includegraphics[scale=0.50]{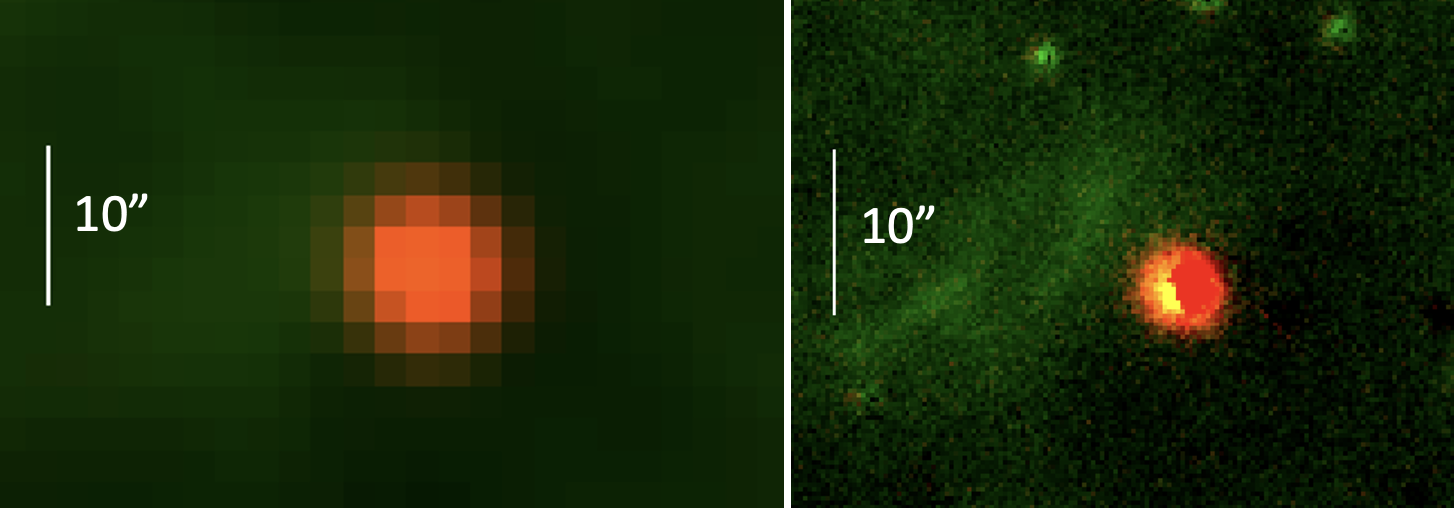}
\caption{\label{PN} LMC planetary nebula SMP83 from field c48 (tile 2).  The MCELS data are shown at left, and DeMCELS at right, with \ha\ in red and \sii\ in green. LMC PNe have very little \sii\ emission and typically appear stellar or nearly stellar.  \citet{shaw06} list this PN as being 3\farcs98 by 3\farcs63 based on HST imagery, and is one of the largest PNe in their survey. }
\end{center}
\end{figure}

\subsection{PNe and other Small Scale Nebulae}

The LMC contains hundreds of identified planetary nebulae (PNe) \citep{reid06,Reid13}, which appear as stellar or nearly stellar sources of \ha\ (and/or \nii) emission since PNe emit at very low levels in \sii. As a practical matter, many of what appear to be stellar residuals in the subtracted \ha\ images in MCELS and DeMCELS surveys are actually unresolved PNe or one of several possible types of emission-line stars in the LMC (W-R stars, X-ray binaries, cataclysmic variables, and the like). \citet{shaw06} obtained images and spectroscopy of a number of LMC PNe using STIS on HST in slitless spectroscopy mode.  In Fig.~\ref{PN} we show the PN SMP83 as seen in both MCELS and DeMCELS. SMP83 is one of the largest PNe listed by \citet{shaw06} with angular size of 3\farcs98 $\times$ 3\farcs63 (0.96 pc$\times$0.87 pc).  Many of the LMC PNe listed by \citet{shaw06} are below 1\arcsec\ in diameter, and hence only marginally-resolved or unresolved even by DeMCELS.

\begin{figure} [hbt!]
\begin{center}
\includegraphics[scale=0.50]{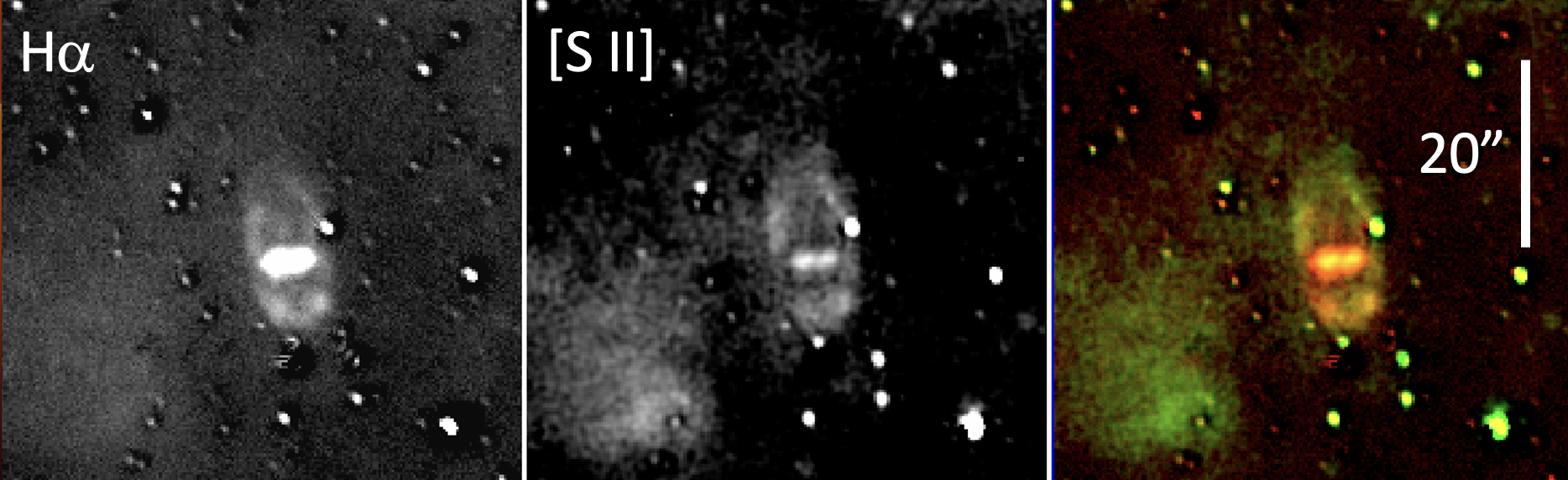}
\caption{\label{Bilobe} DeMCELS data for a small angular size bi-lobed emission nebula from field c42 (tile 7).  The MCELS data only show an unresolved point-like source (primarily from the brighter central edge-on disk) and is not shown here. Otherwise the presentation is the same as in Fig.~\ref{N70}. The bi-lobed structure is $\sim$18\arcsec\ long, much too large for a PN.}
\end{center}
\end{figure}

It is somewhat surprising then that the object shown in Fig.~\ref{Bilobe} has been classified as a PN \citep[PN3464][]{Reid13}. The MCELS data  show only the bright central region as an unresolved small emission region and is not shown in the figure.  
The identification was based on line ratios, since the nebula itself was not resolved in the UK Schmidt survey data used by \citet{Reid13}. Our survey data are apparently the first to resolve the structure of this nebula, showing a bright central region which appears to be an edge-on disk plus a much fainter bi-lobed outer structure.  The extent of the bi-lobed structure is $\sim$18\arcsec\ ($\sim$4.4 pc), which would be exceedingly large for a PN. 
\citet{Reid13} show a spectrum of this source, no doubt dominated by the bright central disk, and unlike most of the \ha\ emission sources, this object shows very strong \nii\ emission, which is indicative of enhanced N abundance. So it is likely that this material is circumstellar medium
that was ejected by the central evolved star or interacting binary, a conclusion supported by the morphology of the nebula.

\begin{figure} [hbt!]
\begin{center}
\includegraphics[scale=0.55]{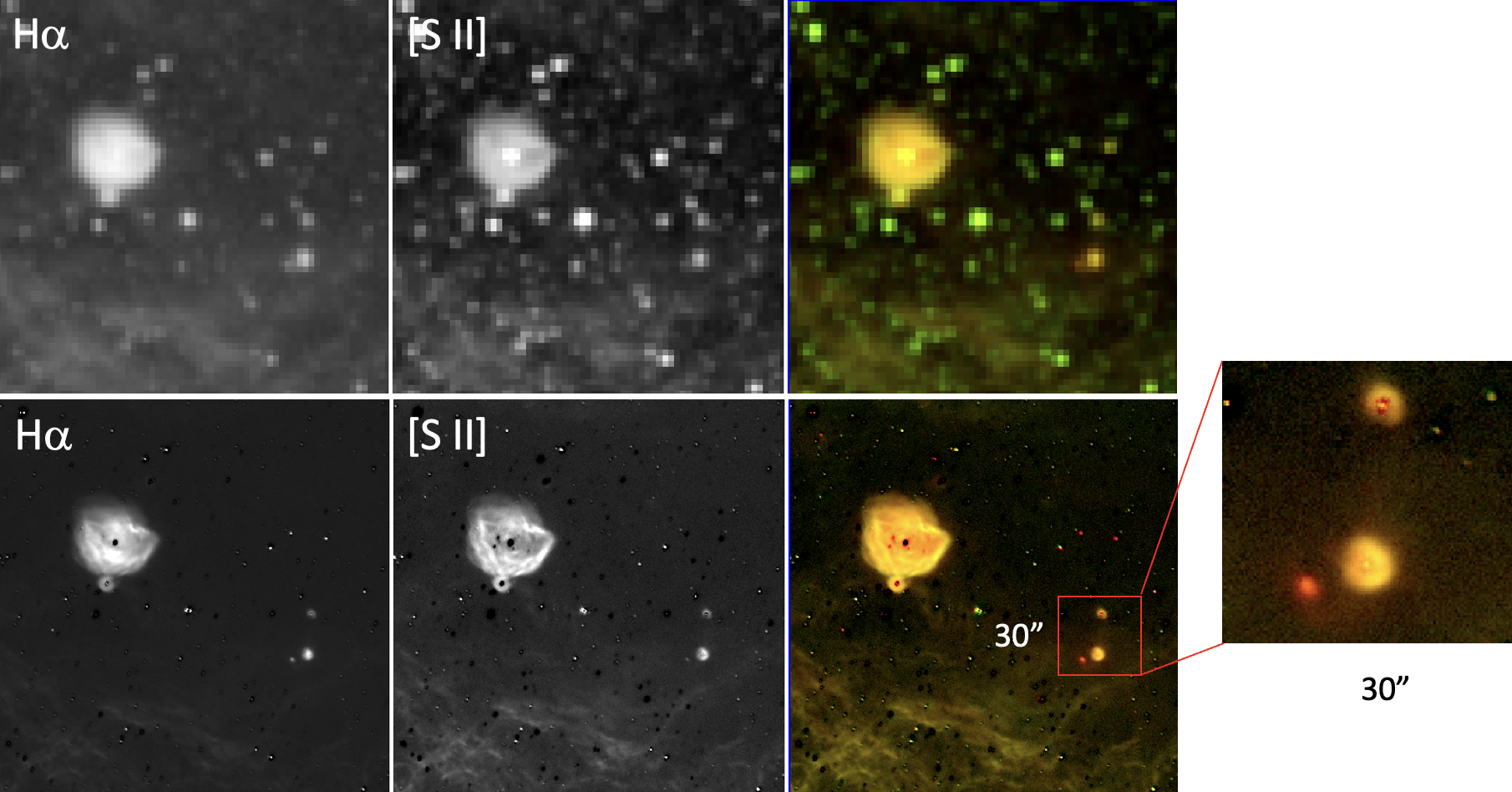}
\caption{\label{SFRs} A comparison of MCELS and DeMCELS for a small region from field c39 that contains several small to very small star forming regions.  Presentation is the same as in Fig.~\ref{N70} except for MCELS, we have used the emission line files prior to continuum subtraction so stars remain visible.   For scale, the box at right is 30\arcsec.  These smallest emission nebulae are totally unresolved in MCELS data.}
\end{center}
\end{figure}

Although it is beyond the scope of this survey overview paper to catalog the numerous small \ha\ nebulae that are unresolved in MCELS but are at least marginally resolved with DECam, we show some examples of such nebulae in 
Fig.~\ref{SFRs}.  This figure highlights a small region north of the N44 complex that contains several small star forming regions of various sizes.  The larger nebula above and left of center is clearly being excited by a small handful of hot interior stars. On the southern edge of this larger nebula, a small ``single star'' nebula can be seen.  The enlargement at right shows a 30\arcsec\ region with three other very small (individual star) nebulae.  Faint structured diffuse emission from the outskirts of the N44 region is also visible in the lower part of the frame.

\subsection{Structured Diffuse Background}

If one ignores the bright regions of emission and looks in the background, the LMC is filled with extensive diffuse emission, some of which has been noted above in earlier figures.  With its larger pixel size, MCELS actually shows just how extensive this emission is, extending outward from and between many of the bright \hii\ region complexes.  Much of this emission is akin to what has been called either the warm ionized medium (WIM) or diffuse ionized gas (DIG) in other nearby galaxies \citep[see][for a review]{haffner09}, but it has received relatively little attention in the LMC. Generally speaking, the DIG shows relatively strong \sii\ emission in comparison to normal photoionized gas, but it is nonetheless photoionized by radiation leaking out from active star forming regions. Some of this faint background emission has a fluffy morphology, and for this component, MCELS actually shows it as effectively or  more so than DeMCELS, owing to the larger pixel scale in the former. However, to the extent that some of this background emission is structured, DeMCELs is effective at showing it.  This is an example of where using the two surveys together can be effective.

\begin{figure} [hbt!]
\begin{center}
\includegraphics[scale=0.55]{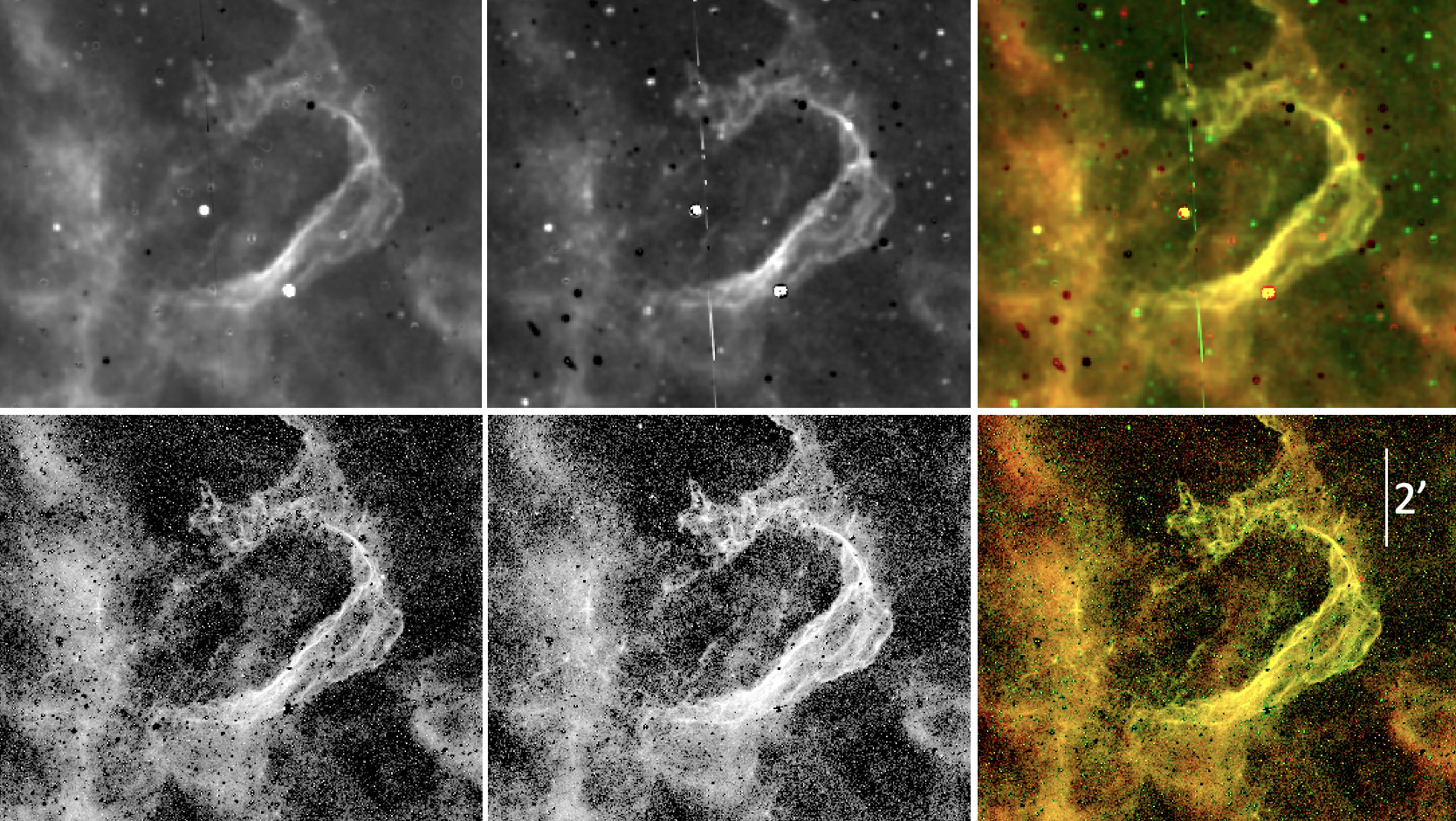}
\caption{\label{DIGregion} A comparison of MCELS and DeMCELS for a region of structured background from field c42 (tile 5). 
Presentation is the same as in Fig.~\ref{N70}.
Although the overall depth of the two surveys is about the same, truly diffuse emission shows up somewhat better in the original MCELS data, owing to its larger pixel size.  However, for any of these background regions that show fine-scale structure, such as the arc of emission shown here, the higher resolution of DeMCELS shows those details better.
}
\end{center}
\end{figure}

Fig.~\ref{DIGregion} shows one example of this comparison, for a background region $\sim0.5^{\circ}$ north of the N44 emission complex. The scaling has been adjusted to enhance what is generally quite faint emission compared with what has been shown in earlier figures. A region of structured emission is seen in projection against more diffuse emission, particularly visible along the left side of the Figure.  Not only does the morphology change between these two components of the background, but the line ratio of \sii\ to \ha\ also changes, being somewhat higher in the structured emission.  As no star cluster or ionizing source is seen within the structured emission and since the \sii/\ha\ ratio is somewhat enhanced compared with other local emission, it is conceivable that this emission represents an old SNR that has nearly faded into the background, but it has not been catalogued as such previously.

\section{Data Availability}\label{sec:datalab}

Our interests in obtaining an improved version of the MCELS survey are primarily related to our interests in the study of shocked gas in SNRs and other settings.  However we also felt that a new survey would be of interest to others for a variety of projects.  As a result, and working with scientists and engineers at NOIRLab, we have made our initial reduction of the DeMCELS survey data available at NOIRLab's 
\href{https://datalab.noirlab.edu/demcels.php}{Data Lab survey page}.
There one can find mechanisms to obtain all or portions of the data.  We expect to acquire additional images of the LMC in the near future and have a companion study of the SMC underway.  We expect future releases of the survey data (including any improvements in the data processing) to be posted on this site as well.

\section{Summary}\label{sec:summary}

We performed a new multiband (\ha\ and \sii) emission-line survey of the LMC using DECam on the Blanco 4-m telescope at CTIO, using  images with the DES r$^\prime$ filter for continuum-subtraction.  Twenty overlapping fields were required to cover the visible extent of the LMC, leading to huge and complex data sets to process into large-area mosaic images.  We have described the data and processing, and then have presented representative examples of various kinds of emission nebulae in the LMC, highlighting the improvements in spatial resolution and diagnostic power provided by the new survey in comparison with the previously available MCELS survey \citep{smith99}.

However, all of the above are simply representative examples.  The power of the DECam survey presented here is that it allows more global studies of a variety of emission nebulae and objects across the entire LMC at a spatial resolution some 3-5 $\times$ higher than the MCELS survey. These data make an ideal complement to other multiwavelength surveys of the LMC \citep[cf.][]{Kim03, meixner06, Maggi16}. 

As a service to the community, we are making the processed data available to enable a much broader range of science than our team's particular scientific interests.  Work on a companion survey of the Small Magellanic Cloud is ongoing.  As these data are processed, we expect to make them available in the same manner. 

\begin{acknowledgments}
This project used data obtained with the Dark Energy Camera (DECam), which was constructed by the Dark Energy Survey (DES) collaboration. Funding for the DES Projects has been provided by the US Department of Energy, the U.S. National Science Foundation, the Ministry of Science and Education of Spain, the Science and Technology Facilities Council of the United Kingdom, the Higher Education Funding Council for England, the National Center for Supercomputing Applications at the University of Illinois at Urbana-Champaign, the Kavli Institute for Cosmological Physics at the University of Chicago, Center for Cosmology and Astro-Particle Physics at the Ohio State University, the Mitchell Institute for Fundamental Physics and Astronomy at Texas A\&M University, Financiadora de Estudos e Projetos, Fundação Carlos Chagas Filho de Amparo à Pesquisa do Estado do Rio de Janeiro, Conselho Nacional de Desenvolvimento Científico e Tecnológico and the Ministério da Ciência, Tecnologia e Inovação, the Deutsche Forschungsgemeinschaft and the Collaborating Institutions in the Dark Energy Survey.

The Collaborating Institutions are Argonne National Laboratory, the University of California at Santa Cruz, the University of Cambridge, Centro de Investigaciones Enérgeticas, Medioambientales y Tecnológicas–Madrid, the University of Chicago, University College London, the DES-Brazil Consortium, the University of Edinburgh, the Eidgenössische Technische Hochschule (ETH) Zürich, Fermi National Accelerator Laboratory, the University of Illinois at Urbana-Champaign, the Institut de Ciències de l’Espai (IEEC/CSIC), the Institut de Física d’Altes Energies, Lawrence Berkeley National Laboratory, the Ludwig-Maximilians Universität München and the associated Excellence Cluster Universe, the University of Michigan, NSF NOIRLab, the University of Nottingham, the Ohio State University, the OzDES Membership Consortium, the University of Pennsylvania, the University of Portsmouth, SLAC National Accelerator Laboratory, Stanford University, the University of Sussex, and Texas A\&M University.

Based on observations at NSF Cerro Tololo Inter-American Observatory, NSF NOIRLab (NOIRLab Prop. ID 2018A-0909; PI: T. Puzia; NOIRLab Prop. ID 2018B-0908; PI: T. Puzia; and NOIRLab PropID 2021B-0060; PI: S. Points), which is managed by the Association of Universities for Research in Astronomy (AURA) under a cooperative agreement with the U.S. National Science Foundation.

This work has made use of data from the European Space Agency (ESA)
mission Gaia (\url{https://www.cosmos.esa.int/gaia}), processed by
the Gaia Data Processing and Analysis Consortium (DPAC,
\url{https://www.cosmos.esa.int/web/gaia/dpac/consortium}). Funding
for the DPAC has been provided by national institutions, in particular
the institutions participating in the Gaia Multilateral Agreement.

SDP would like to thank Thomas Puzia and Eric Peng for the purchase of the N662 filter and starting this work with their \ha\ survey of the LMC.  SDP would also like to thank Kyoungsoo Lee and the ODIN survey for permitting us to use the N673 filter in this investigation.
WPB acknowledges support from the Johns Hopkins Center for Astronomical Sciences during this work.  PFW acknowledges the support of the NSF through grant AST-1714281.
YHC acknowledges the support of the grants NSTC 112-2112-M-001-065
and NSTC 111-2112-M-001-063 from the National Science and Technology Council of Taiwan.
RMW would like to acknowledge the contributions of the following Columbus State University undergraduate students for data review and object searches: Kayleen Linge, Devin Janeway, Sharmaine Motin, A'naja Houston, Delta Flowers, Griffin McLeroy, Cory Mitchell, Trinity Smith, Samuel Kimball, and William Morgan.

\end{acknowledgments}

\facility{Blanco}

\software{SWarp \citep{SWARP}, Astropy \citep{astropy:2013, astropy:2018, astropy:2022}, Matplotlib \citep{Hunter:2007}, NumPy \citep{harris2020array}, SciPy \citep{2020SciPy-NMeth}, and SAOImage DS9 \citep{ds9}}

\bibliographystyle{aasjournal}
\bibliography{mc_ref}

\begin{thebibliography}{}
\expandafter\ifx\csname natexlab\endcsname\relax\def\natexlab#1{#1}\fi
\providecommand{\url}[1]{\href{#1}{#1}}
\providecommand{\dodoi}[1]{doi:~\href{http://doi.org/#1}{\nolinkurl{#1}}}
\providecommand{\doeprint}[1]{\href{http://ascl.net/#1}{\nolinkurl{http://ascl.net/#1}}}
\providecommand{\doarXiv}[1]{\href{https://arxiv.org/abs/#1}{\nolinkurl{https://arxiv.org/abs/#1}}}

\bibitem[{{Astropy Collaboration} {et~al.}(2013){Astropy Collaboration},
  {Robitaille}, {Tollerud}, {Greenfield}, {Droettboom}, {Bray}, {Aldcroft},
  {Davis}, {Ginsburg}, {Price-Whelan}, {Kerzendorf}, {Conley}, {Crighton},
  {Barbary}, {Muna}, {Ferguson}, {Grollier}, {Parikh}, {Nair}, {Unther},
  {Deil}, {Woillez}, {Conseil}, {Kramer}, {Turner}, {Singer}, {Fox}, {Weaver},
  {Zabalza}, {Edwards}, {Azalee Bostroem}, {Burke}, {Casey}, {Crawford},
  {Dencheva}, {Ely}, {Jenness}, {Labrie}, {Lim}, {Pierfederici}, {Pontzen},
  {Ptak}, {Refsdal}, {Servillat}, \& {Streicher}}]{astropy:2013}
{Astropy Collaboration}, {Robitaille}, T.~P., {Tollerud}, E.~J., {et~al.} 2013,
  \aap, 558, A33, \dodoi{10.1051/0004-6361/201322068}

\bibitem[{{Astropy Collaboration} {et~al.}(2018){Astropy Collaboration},
  {Price-Whelan}, {Sip{\H{o}}cz}, {G{\"u}nther}, {Lim}, {Crawford}, {Conseil},
  {Shupe}, {Craig}, {Dencheva}, {Ginsburg}, {Vand erPlas}, {Bradley},
  {P{\'e}rez-Su{\'a}rez}, {de Val-Borro}, {Aldcroft}, {Cruz}, {Robitaille},
  {Tollerud}, {Ardelean}, {Babej}, {Bach}, {Bachetti}, {Bakanov}, {Bamford},
  {Barentsen}, {Barmby}, {Baumbach}, {Berry}, {Biscani}, {Boquien}, {Bostroem},
  {Bouma}, {Brammer}, {Bray}, {Breytenbach}, {Buddelmeijer}, {Burke},
  {Calderone}, {Cano Rodr{\'\i}guez}, {Cara}, {Cardoso}, {Cheedella}, {Copin},
  {Corrales}, {Crichton}, {D'Avella}, {Deil}, {Depagne}, {Dietrich}, {Donath},
  {Droettboom}, {Earl}, {Erben}, {Fabbro}, {Ferreira}, {Finethy}, {Fox},
  {Garrison}, {Gibbons}, {Goldstein}, {Gommers}, {Greco}, {Greenfield},
  {Groener}, {Grollier}, {Hagen}, {Hirst}, {Homeier}, {Horton}, {Hosseinzadeh},
  {Hu}, {Hunkeler}, {Ivezi{\'c}}, {Jain}, {Jenness}, {Kanarek}, {Kendrew},
  {Kern}, {Kerzendorf}, {Khvalko}, {King}, {Kirkby}, {Kulkarni}, {Kumar},
  {Lee}, {Lenz}, {Littlefair}, {Ma}, {Macleod}, {Mastropietro}, {McCully},
  {Montagnac}, {Morris}, {Mueller}, {Mumford}, {Muna}, {Murphy}, {Nelson},
  {Nguyen}, {Ninan}, {N{\"o}the}, {Ogaz}, {Oh}, {Parejko}, {Parley}, {Pascual},
  {Patil}, {Patil}, {Plunkett}, {Prochaska}, {Rastogi}, {Reddy Janga},
  {Sabater}, {Sakurikar}, {Seifert}, {Sherbert}, {Sherwood-Taylor}, {Shih},
  {Sick}, {Silbiger}, {Singanamalla}, {Singer}, {Sladen}, {Sooley},
  {Sornarajah}, {Streicher}, {Teuben}, {Thomas}, {Tremblay}, {Turner},
  {Terr{\'o}n}, {van Kerkwijk}, {de la Vega}, {Watkins}, {Weaver}, {Whitmore},
  {Woillez}, {Zabalza}, \& {Astropy Contributors}}]{astropy:2018}
{Astropy Collaboration}, {Price-Whelan}, A.~M., {Sip{\H{o}}cz}, B.~M., {et~al.}
  2018, \aj, 156, 123, \dodoi{10.3847/1538-3881/aabc4f}

\bibitem[{{Astropy Collaboration} {et~al.}(2022){Astropy Collaboration},
  {Price-Whelan}, {Lim}, {Earl}, {Starkman}, {Bradley}, {Shupe}, {Patil},
  {Corrales}, {Brasseur}, {N{"o}the}, {Donath}, {Tollerud}, {Morris},
  {Ginsburg}, {Vaher}, {Weaver}, {Tocknell}, {Jamieson}, {van Kerkwijk},
  {Robitaille}, {Merry}, {Bachetti}, {G{"u}nther}, {Aldcroft},
  {Alvarado-Montes}, {Archibald}, {B{'o}di}, {Bapat}, {Barentsen}, {Baz{'a}n},
  {Biswas}, {Boquien}, {Burke}, {Cara}, {Cara}, {Conroy}, {Conseil}, {Craig},
  {Cross}, {Cruz}, {D'Eugenio}, {Dencheva}, {Devillepoix}, {Dietrich},
  {Eigenbrot}, {Erben}, {Ferreira}, {Foreman-Mackey}, {Fox}, {Freij}, {Garg},
  {Geda}, {Glattly}, {Gondhalekar}, {Gordon}, {Grant}, {Greenfield}, {Groener},
  {Guest}, {Gurovich}, {Handberg}, {Hart}, {Hatfield-Dodds}, {Homeier},
  {Hosseinzadeh}, {Jenness}, {Jones}, {Joseph}, {Kalmbach}, {Karamehmetoglu},
  {Ka{l}uszy{'n}ski}, {Kelley}, {Kern}, {Kerzendorf}, {Koch}, {Kulumani},
  {Lee}, {Ly}, {Ma}, {MacBride}, {Maljaars}, {Muna}, {Murphy}, {Norman},
  {O'Steen}, {Oman}, {Pacifici}, {Pascual}, {Pascual-Granado}, {Patil},
  {Perren}, {Pickering}, {Rastogi}, {Roulston}, {Ryan}, {Rykoff}, {Sabater},
  {Sakurikar}, {Salgado}, {Sanghi}, {Saunders}, {Savchenko}, {Schwardt},
  {Seifert-Eckert}, {Shih}, {Jain}, {Shukla}, {Sick}, {Simpson},
  {Singanamalla}, {Singer}, {Singhal}, {Sinha}, {Sip{H{o}}cz}, {Spitler},
  {Stansby}, {Streicher}, {{{S}}umak}, {Swinbank}, {Taranu}, {Tewary},
  {Tremblay}, {Val-Borro}, {Van Kooten}, {Vasovi{'c}}, {Verma}, {de Miranda
  Cardoso}, {Williams}, {Wilson}, {Winkel}, {Wood-Vasey}, {Xue}, {Yoachim},
  {Zhang}, {Zonca}, \& {Astropy Project Contributors}}]{astropy:2022}
{Astropy Collaboration}, {Price-Whelan}, A.~M., {Lim}, P.~L., {et~al.} 2022,
  \apj, 935, 167, \dodoi{10.3847/1538-4357/ac7c74}

\bibitem[{{Bertin} {et~al.}(2002){Bertin}, {Mellier}, {Radovich}, {Missonnier},
  {Didelon}, \& {Morin}}]{SWARP}
{Bertin}, E., {Mellier}, Y., {Radovich}, M., {et~al.} 2002, in Astronomical
  Society of the Pacific Conference Series, Vol. 281, Astronomical Data
  Analysis Software and Systems XI, ed. D.~A. {Bohlender}, D.~{Durand}, \&
  T.~H. {Handley}, 228

\bibitem[{{Blair} {et~al.}(2006){Blair}, {Ghavamian}, {Sankrit}, \&
  {Danforth}}]{blair06}
{Blair}, W.~P., {Ghavamian}, P., {Sankrit}, R., \& {Danforth}, C.~W. 2006,
  \apjs, 165, 480, \dodoi{10.1086/505346}

\bibitem[{{Bozzetto} {et~al.}(2017){Bozzetto}, {Filipovi{\'c}}, {Vukoti{\'c}},
  {Pavlovi{\'c}}, {Uro{\v s}evi{\'c}}, {Kavanagh}, {Arbutina}, {Maggi},
  {Sasaki}, {Haberl}, {Crawford}, {Roper}, {Grieve}, \& {Points}}]{bozzetto17}
{Bozzetto}, L.~M., {Filipovi{\'c}}, M.~D., {Vukoti{\'c}}, B., {et~al.} 2017,
  \apjs, 230, 2, \dodoi{10.3847/1538-4365/aa653c}

\bibitem[{{Bozzetto} {et~al.}(2023){Bozzetto}, {Filipovi{\'c}}, {Sano},
  {Alsaberi}, {Barnes}, {Boji{\v{c}}i{\'c}}, {Brose}, {Chomiuk}, {Crawford},
  {Dai}, {Ghavam}, {Haberl}, {Hill}, {Hopkins}, {Ingallinera}, {Jarrett},
  {Kavanagh}, {Koribalski}, {Kothes}, {Leahy}, {Lenc}, {Leonidaki}, {Maggi},
  {Maitra}, {Matthew}, {Payne}, {Pennock}, {Points}, {Reid}, {Riggi}, {Rowell},
  {Sasaki}, {Safi-Harb}, {van Loon}, {Tothill}, {Uro{\v{s}}evi{\'c}}, \&
  {Zangrandi}}]{bozzetto23}
{Bozzetto}, L.~M., {Filipovi{\'c}}, M.~D., {Sano}, H., {et~al.} 2023, \mnras,
  518, 2574, \dodoi{10.1093/mnras/stac2922}

\bibitem[{{Carlos Reyes} {et~al.}(2015){Carlos Reyes}, {Reyes Navarro},
  {Mel{\'e}ndez}, {Steiner}, \& {Elizalde}}]{CarlosReyes15}
{Carlos Reyes}, R.~E., {Reyes Navarro}, F.~A., {Mel{\'e}ndez}, J., {Steiner},
  J., \& {Elizalde}, F. 2015, \rmxaa, 51, 135

\bibitem[{Choudhury {et~al.}(2021)Choudhury, de Grijs, Bekki, Cioni, Ivanov,
  van Loon, Miller, Niederhofer, Oliveira, Ripepi, Sun, \&
  Subramanian}]{Choudhury2021}
Choudhury, S., de Grijs, R., Bekki, K., {et~al.} 2021, Monthly Notices of the
  Royal Astronomical Society, 507, 4752, \dodoi{10.1093/mnras/stab2446}

\bibitem[{{Chu} {et~al.}(1995){Chu}, {Dickel}, {Staveley-Smith}, {Osterberg},
  \& {Smith}}]{Chu95}
{Chu}, Y.-H., {Dickel}, J.~R., {Staveley-Smith}, L., {Osterberg}, J., \&
  {Smith}, R.~C. 1995, \aj, 109, 1729, \dodoi{10.1086/117401}

\bibitem[{{Chu} \& {Kennicutt}(1988)}]{Chu88a}
{Chu}, Y.-H., \& {Kennicutt}, Robert~C., J. 1988, \aj, 95, 1111,
  \dodoi{10.1086/114706}

\bibitem[{{Chu} \& {Mac Low}(1990)}]{Chu90}
{Chu}, Y.-H., \& {Mac Low}, M.-M. 1990, \apj, 365, 510, \dodoi{10.1086/169505}

\bibitem[{{Chu} {et~al.}(1993){Chu}, {Mac Low}, {Garcia-Segura}, {Wakker}, \&
  {Kennicutt}}]{Chu1993}
{Chu}, Y.-H., {Mac Low}, M.-M., {Garcia-Segura}, G., {Wakker}, B., \&
  {Kennicutt}, Robert~C., J. 1993, \apj, 414, 213, \dodoi{10.1086/173069}

\bibitem[{{Collischon} {et~al.}(2021){Collischon}, {Sasaki}, {Mecke}, {Points},
  \& {Klatt}}]{Collischon21}
{Collischon}, C., {Sasaki}, M., {Mecke}, K., {Points}, S.~D., \& {Klatt}, M.~A.
  2021, \aap, 653, A16, \dodoi{10.1051/0004-6361/202040153}

\bibitem[{{Crawford} {et~al.}(2010){Crawford}, {Filipovi{\'c}}, {Haberl},
  {Pietsch}, {Payne}, \& {de Horta}}]{Crawford10}
{Crawford}, E.~J., {Filipovi{\'c}}, M.~D., {Haberl}, F., {et~al.} 2010, \aap,
  518, A35, \dodoi{10.1051/0004-6361/201014767}

\bibitem[{{Danforth} \& {Blair}(2006)}]{Danforth06}
{Danforth}, C.~W., \& {Blair}, W.~P. 2006, \apj, 646, 205,
  \dodoi{10.1086/504706}

\bibitem[{{Davies} {et~al.}(1976){Davies}, {Elliott}, \& {Meaburn}}]{Davies76}
{Davies}, R.~D., {Elliott}, K.~H., \& {Meaburn}, J. 1976, \memras, 81, 89

\bibitem[{{de Horta} {et~al.}(2012){de Horta}, {Filipovi{\'c}}, {Bozzetto},
  {Maggi}, {Haberl}, {Crawford}, {Sasaki}, {Uro{\v{s}}evi{\'c}}, {Pietsch},
  {Gruendl}, {Dickel}, {Tothill}, {Chu}, {Payne}, \& {Collier}}]{deHorta12}
{de Horta}, A.~Y., {Filipovi{\'c}}, M.~D., {Bozzetto}, L.~M., {et~al.} 2012,
  \aap, 540, A25, \dodoi{10.1051/0004-6361/201118694}

\bibitem[{{Dopita} {et~al.}(1981){Dopita}, {Ford}, {McGregor}, {Mathewson}, \&
  {Wilson}}]{Dopita81}
{Dopita}, M.~A., {Ford}, V.~L., {McGregor}, P.~J., {Mathewson}, D.~S., \&
  {Wilson}, I.~R. 1981, \apj, 250, 103, \dodoi{10.1086/159352}

\bibitem[{{Dufour}(1975)}]{Dufour75}
{Dufour}, R.~J. 1975, \apj, 195, 315, \dodoi{10.1086/153330}

\bibitem[{{Dunne} {et~al.}(2001){Dunne}, {Points}, \& {Chu}}]{dunne01}
{Dunne}, B.~C., {Points}, S.~D., \& {Chu}, Y.-H. 2001, \apjs, 136, 119,
  \dodoi{10.1086/321794}

\bibitem[{{Flaugher} {et~al.}(2015){Flaugher}, {Diehl}, {Honscheid}, {Abbott},
  {Alvarez}, {Angstadt}, {Annis}, {Antonik}, {Ballester}, {Beaufore},
  {Bernstein}, {Bernstein}, {Bigelow}, {Bonati}, {Boprie}, {Brooks},
  {Buckley-Geer}, {Campa}, {Cardiel-Sas}, {Castander}, {Castilla}, {Cease},
  {Cela-Ruiz}, {Chappa}, {Chi}, {Cooper}, {da Costa}, {Dede}, {Derylo},
  {DePoy}, {de Vicente}, {Doel}, {Drlica-Wagner}, {Eiting}, {Elliott}, {Emes},
  {Estrada}, {Fausti Neto}, {Finley}, {Flores}, {Frieman}, {Gerdes},
  {Gladders}, {Gregory}, {Gutierrez}, {Hao}, {Holland}, {Holm}, {Huffman},
  {Jackson}, {James}, {Jonas}, {Karcher}, {Karliner}, {Kent}, {Kessler},
  {Kozlovsky}, {Kron}, {Kubik}, {Kuehn}, {Kuhlmann}, {Kuk}, {Lahav}, {Lathrop},
  {Lee}, {Levi}, {Lewis}, {Li}, {Mandrichenko}, {Marshall}, {Martinez},
  {Merritt}, {Miquel}, {Mu{\~n}oz}, {Neilsen}, {Nichol}, {Nord}, {Ogando},
  {Olsen}, {Palaio}, {Patton}, {Peoples}, {Plazas}, {Rauch}, {Reil}, {Rheault},
  {Roe}, {Rogers}, {Roodman}, {Sanchez}, {Scarpine}, {Schindler}, {Schmidt},
  {Schmitt}, {Schubnell}, {Schultz}, {Schurter}, {Scott}, {Serrano}, {Shaw},
  {Smith}, {Soares-Santos}, {Stefanik}, {Stuermer}, {Suchyta}, {Sypniewski},
  {Tarle}, {Thaler}, {Tighe}, {Tran}, {Tucker}, {Walker}, {Wang}, {Watson},
  {Weaverdyck}, {Wester}, {Woods}, {Yanny}, \& {DES
  Collaboration}}]{Flaugher15}
{Flaugher}, B., {Diehl}, H.~T., {Honscheid}, K., {et~al.} 2015, \aj, 150, 150,
  \dodoi{10.1088/0004-6256/150/5/150}

\bibitem[{{Haffner} {et~al.}(2009){Haffner}, {Dettmar}, {Beckman}, {Wood},
  {Slavin}, {Giammanco}, {Madsen}, {Zurita}, \& {Reynolds}}]{haffner09}
{Haffner}, L.~M., {Dettmar}, R.~J., {Beckman}, J.~E., {et~al.} 2009, Reviews of
  Modern Physics, 81, 969, \dodoi{10.1103/RevModPhys.81.969}

\bibitem[{Harris {et~al.}(2020)Harris, Millman, van~der Walt, Gommers,
  Virtanen, Cournapeau, Wieser, Taylor, Berg, Smith, Kern, Picus, Hoyer, van
  Kerkwijk, Brett, Haldane, del R{\'{i}}o, Wiebe, Peterson,
  G{\'{e}}rard-Marchant, Sheppard, Reddy, Weckesser, Abbasi, Gohlke, \&
  Oliphant}]{harris2020array}
Harris, C.~R., Millman, K.~J., van~der Walt, S.~J., {et~al.} 2020, Nature, 585,
  357, \dodoi{10.1038/s41586-020-2649-2}

\bibitem[{{Harris} \& {Zaritsky}(2009)}]{HZ2009}
{Harris}, J., \& {Zaritsky}, D. 2009, \aj, 138, 1243,
  \dodoi{10.1088/0004-6256/138/5/1243}

\bibitem[{{Hassani} {et~al.}(2022){Hassani}, {Tabatabaei}, {Hughes},
  {Chastenet}, {McLeod}, {Schinnerer}, \& {Nasiri}}]{Hassani22}
{Hassani}, H., {Tabatabaei}, F., {Hughes}, A., {et~al.} 2022, \mnras, 510, 11,
  \dodoi{10.1093/mnras/stab3202}

\bibitem[{{Henize}(1956)}]{Henize56}
{Henize}, K.~G. 1956, \apjs, 2, 315, \dodoi{10.1086/190025}

\bibitem[{{Honscheid} \& {DePoy}(2008)}]{Honscheid08}
{Honscheid}, K., \& {DePoy}, D.~L. 2008, arXiv e-prints, arXiv:0810.3600,
  \dodoi{10.48550/arXiv.0810.3600}

\bibitem[{{Hung} {et~al.}(2021){Hung}, {Ou}, {Chu}, {Gruendl}, \&
  {Li}}]{Hung21}
{Hung}, C.~S., {Ou}, P.-S., {Chu}, Y.-H., {Gruendl}, R.~A., \& {Li}, C.-J.
  2021, \apjs, 252, 21, \dodoi{10.3847/1538-4365/abcc00}

\bibitem[{Hunter(2007)}]{Hunter:2007}
Hunter, J.~D. 2007, Computing in Science \& Engineering, 9, 90,
  \dodoi{10.1109/MCSE.2007.55}

\bibitem[{{Joye} \& {Mandel}(2003)}]{ds9}
{Joye}, W.~A., \& {Mandel}, E. 2003, in Astronomical Society of the Pacific
  Conference Series, Vol. 295, Astronomical Data Analysis Software and Systems
  XII, ed. H.~E. {Payne}, R.~I. {Jedrzejewski}, \& R.~N. {Hook}, 489

\bibitem[{{Kavanagh}(2020)}]{Kavanagh20}
{Kavanagh}, P.~J. 2020, \apss, 365, 6, \dodoi{10.1007/s10509-019-3719-5}

\bibitem[{{Kavanagh} {et~al.}(2015{\natexlab{a}}){Kavanagh}, {Sasaki},
  {Bozzetto}, {Filipovi{\'c}}, {Points}, {Maggi}, \& {Haberl}}]{Kavanagh15a}
{Kavanagh}, P.~J., {Sasaki}, M., {Bozzetto}, L.~M., {et~al.}
  2015{\natexlab{a}}, \aap, 573, A73, \dodoi{10.1051/0004-6361/201424354}

\bibitem[{{Kavanagh} {et~al.}(2015{\natexlab{b}}){Kavanagh}, {Sasaki},
  {Bozzetto}, {Points}, {Filipovi{\'c}}, {Maggi}, {Haberl}, \&
  {Crawford}}]{Kavanagh15b}
---. 2015{\natexlab{b}}, \aap, 583, A121, \dodoi{10.1051/0004-6361/201526987}

\bibitem[{{Kavanagh} {et~al.}(2022){Kavanagh}, {Sasaki}, {Filipovi{\'c}},
  {Points}, {Bozzetto}, {Haberl}, {Maggi}, \& {Maitra}}]{Kavanagh22}
{Kavanagh}, P.~J., {Sasaki}, M., {Filipovi{\'c}}, M.~D., {et~al.} 2022, \mnras,
  515, 4099, \dodoi{10.1093/mnras/stac813}

\bibitem[{{Kim} {et~al.}(2003){Kim}, {Staveley-Smith}, {Dopita}, {Sault},
  {Freeman}, {Lee}, \& {Chu}}]{Kim03}
{Kim}, S., {Staveley-Smith}, L., {Dopita}, M.~A., {et~al.} 2003, \apjs, 148,
  473, \dodoi{10.1086/376980}

\bibitem[{{Long} {et~al.}(2022){Long}, {Blair}, {Winkler}, {Della Bruna},
  {Adamo}, {McLeod}, \& {Amram}}]{long22}
{Long}, K.~S., {Blair}, W.~P., {Winkler}, P.~F., {et~al.} 2022, \apj, 929, 144,
  \dodoi{10.3847/1538-4357/ac5aa3}

\bibitem[{{Lucke} \& {Hodge}(1970)}]{Lucke70}
{Lucke}, P.~B., \& {Hodge}, P.~W. 1970, \aj, 75, 171, \dodoi{10.1086/110959}

\bibitem[{{Maggi} {et~al.}(2016){Maggi}, {Haberl}, {Kavanagh}, {Sasaki},
  {Bozzetto}, {Filipovi{\'c}}, {Vasilopoulos}, {Pietsch}, {Points}, {Chu},
  {Dickel}, {Ehle}, {Williams}, \& {Greiner}}]{Maggi16}
{Maggi}, P., {Haberl}, F., {Kavanagh}, P.~J., {et~al.} 2016, \aap, 585, A162,
  \dodoi{10.1051/0004-6361/201526932}

\bibitem[{{Mathewson} {et~al.}(1983){Mathewson}, {Ford}, {Dopita}, {Tuohy},
  {Long}, \& {Helfand}}]{mathewson83}
{Mathewson}, D.~S., {Ford}, V.~L., {Dopita}, M.~A., {et~al.} 1983, \apjs, 51,
  345, \dodoi{10.1086/190854}

\bibitem[{{Meaburn} {et~al.}(2010){Meaburn}, {Redman}, {Boumis}, \&
  {Harvey}}]{Meaburn10}
{Meaburn}, J., {Redman}, M.~P., {Boumis}, P., \& {Harvey}, E. 2010, \mnras,
  408, 1249, \dodoi{10.1111/j.1365-2966.2010.17204.x}

\bibitem[{{Meaburn} {et~al.}(1993){Meaburn}, {Wang}, {Palmer}, \&
  {Lopez}}]{Meaburn93}
{Meaburn}, J., {Wang}, L., {Palmer}, J., \& {Lopez}, J.~A. 1993, \mnras, 263,
  L6, \dodoi{10.1093/mnras/263.1.L6}

\bibitem[{{Meixner} {et~al.}(2006){Meixner}, {Gordon}, {Indebetouw}, {Hora},
  {Whitney}, {Blum}, {Reach}, {Bernard}, {Meade}, {Babler}, {Engelbracht},
  {For}, {Misselt}, {Vijh}, {Leitherer}, {Cohen}, {Churchwell}, {Boulanger},
  {Frogel}, {Fukui}, {Gallagher}, {Gorjian}, {Harris}, {Kelly}, {Kawamura},
  {Kim}, {Latter}, {Madden}, {Markwick-Kemper}, {Mizuno}, {Mizuno}, {Mould},
  {Nota}, {Oey}, {Olsen}, {Onishi}, {Paladini}, {Panagia}, {Perez-Gonzalez},
  {Shibai}, {Sato}, {Smith}, {Staveley-Smith}, {Tielens}, {Ueta}, {van Dyk},
  {Volk}, {Werner}, \& {Zaritsky}}]{meixner06}
{Meixner}, M., {Gordon}, K.~D., {Indebetouw}, R., {et~al.} 2006, \aj, 132,
  2268, \dodoi{10.1086/508185}

\bibitem[{{Oey}(1996{\natexlab{a}})}]{Oey96a}
{Oey}, M.~S. 1996{\natexlab{a}}, \apjs, 104, 71, \dodoi{10.1086/192292}

\bibitem[{{Oey}(1996{\natexlab{b}})}]{Oey96b}
---. 1996{\natexlab{b}}, \apj, 465, 231, \dodoi{10.1086/177415}

\bibitem[{{Oey}(1996{\natexlab{c}})}]{Oey96c}
---. 1996{\natexlab{c}}, \apj, 467, 666, \dodoi{10.1086/177642}

\bibitem[{{Oey} {et~al.}(2000){Oey}, {Dopita}, {Shields}, \& {Smith}}]{Oey00}
{Oey}, M.~S., {Dopita}, M.~A., {Shields}, J.~C., \& {Smith}, R.~C. 2000, \apjs,
  128, 511, \dodoi{10.1086/313396}

\bibitem[{{Oey} {et~al.}(2002){Oey}, {Groves}, {Staveley-Smith}, \&
  {Smith}}]{Oey02}
{Oey}, M.~S., {Groves}, B., {Staveley-Smith}, L., \& {Smith}, R.~C. 2002, \aj,
  123, 255, \dodoi{10.1086/338092}

\bibitem[{{Pellegrini} {et~al.}(2012){Pellegrini}, {Oey}, {Winkler}, {Points},
  {Smith}, {Jaskot}, \& {Zastrow}}]{Pellegrini12}
{Pellegrini}, E.~W., {Oey}, M.~S., {Winkler}, P.~F., {et~al.} 2012, \apj, 755,
  40, \dodoi{10.1088/0004-637X/755/1/40}

\bibitem[{{Pietrzy{\'n}ski} {et~al.}(2019){Pietrzy{\'n}ski}, {Graczyk},
  {Gallenne}, {Gieren}, {Thompson}, {Pilecki}, {Karczmarek}, {G{\'o}rski},
  {Suchomska}, {Taormina}, {Zgirski}, {Wielg{\'o}rski}, {Ko{\l}aczkowski},
  {Konorski}, {Villanova}, {Nardetto}, {Kervella}, {Bresolin}, {Kudritzki},
  {Storm}, {Smolec}, \& {Narloch}}]{Pietrzynski19}
{Pietrzy{\'n}ski}, G., {Graczyk}, D., {Gallenne}, A., {et~al.} 2019, \nat, 567,
  200, \dodoi{10.1038/s41586-019-0999-4}

\bibitem[{{Reid} \& {Parker}(2006)}]{reid06}
{Reid}, W.~A., \& {Parker}, Q.~A. 2006, \mnras, 373, 521,
  \dodoi{10.1111/j.1365-2966.2006.11087.x}

\bibitem[{{Reid} \& {Parker}(2013)}]{Reid13}
---. 2013, \mnras, 436, 604, \dodoi{10.1093/mnras/stt1609}

\bibitem[{{Sasaki} {et~al.}(2022){Sasaki}, {Knies}, {Haberl}, {Maitra}, {Kerp},
  {Bykov}, {Dennerl}, {Filipovi{\'c}}, {Freyberg}, {Koribalski}, {Points}, \&
  {Staveley-Smith}}]{Sasaki22}
{Sasaki}, M., {Knies}, J., {Haberl}, F., {et~al.} 2022, \aap, 661, A37,
  \dodoi{10.1051/0004-6361/202141054}

\bibitem[{{Shaw} {et~al.}(2006){Shaw}, {Stanghellini}, {Villaver}, \&
  {Mutchler}}]{shaw06}
{Shaw}, R.~A., {Stanghellini}, L., {Villaver}, E., \& {Mutchler}, M. 2006,
  \apjs, 167, 201, \dodoi{10.1086/508469}

\bibitem[{Shipp {et~al.}(2021)Shipp, Erkal, Drlica-Wagner, Li, Pace, Koposov,
  Cullinane, Costa, Ji, Kuehn, Lewis, Mackey, Simpson, Wan, Zucker,
  Bland-Hawthorn, Ferguson, Lilleengen, \& Collaboration)}]{Shipp_2021}
Shipp, N., Erkal, D., Drlica-Wagner, A., {et~al.} 2021, The Astrophysical
  Journal, 923, 149, \dodoi{10.3847/1538-4357/ac2e93}

\bibitem[{{Skelton} {et~al.}(1999){Skelton}, {Waller}, {Gelderman}, {Brown},
  {Woodgate}, {Caulet}, \& {Schommer}}]{Skelton99}
{Skelton}, B.~P., {Waller}, W.~H., {Gelderman}, R.~F., {et~al.} 1999, \pasp,
  111, 465, \dodoi{10.1086/316346}

\bibitem[{{Smith} \& {MCELS Team}(1999)}]{smith99}
{Smith}, R.~C., \& {MCELS Team}. 1999, in IAU Symposium, Vol. 190, New Views of
  the Magellanic Clouds, ed. Y.-H. {Chu}, N.~{Suntzeff}, J.~{Hesser}, \&
  D.~{Bohlender}, 28

\bibitem[{{Valdes} {et~al.}(2014){Valdes}, {Gruendl}, \& {DES
  Project}}]{DECamCP}
{Valdes}, F., {Gruendl}, R., \& {DES Project}. 2014, in Astronomical Society of
  the Pacific Conference Series, Vol. 485, Astronomical Data Analysis Software
  and Systems XXIII, ed. N.~{Manset} \& P.~{Forshay}, 379

\bibitem[{Virtanen {et~al.}(2020)Virtanen, Gommers, Oliphant, Haberland, Reddy,
  Cournapeau, Burovski, Peterson, Weckesser, Bright, {van der Walt}, Brett,
  Wilson, Millman, Mayorov, Nelson, Jones, Kern, Larson, Carey, Polat, Feng,
  Moore, {VanderPlas}, Laxalde, Perktold, Cimrman, Henriksen, Quintero, Harris,
  Archibald, Ribeiro, Pedregosa, {van Mulbregt}, \& {SciPy 1.0
  Contributors}}]{2020SciPy-NMeth}
Virtanen, P., Gommers, R., Oliphant, T.~E., {et~al.} 2020, Nature Methods, 17,
  261, \dodoi{10.1038/s41592-019-0686-2}

\bibitem[{{Wang}(1992)}]{Wang92}
{Wang}, L. 1992, The Messenger, 69, 34

\bibitem[{{Wang} \& {Helfand}(1991)}]{wang91a}
{Wang}, Q., \& {Helfand}, D.~J. 1991, \apj, 373, 497, \dodoi{10.1086/170069}

\bibitem[{{Warth} {et~al.}(2014){Warth}, {Sasaki}, {Kavanagh}, {Filipovi{\'c}},
  {Points}, \& {Bozzetto}}]{Warth14}
{Warth}, G., {Sasaki}, M., {Kavanagh}, P.~J., {et~al.} 2014, \aap, 567, A136,
  \dodoi{10.1051/0004-6361/201423575}

\bibitem[{{Williams}(2009)}]{Williams2009}
{Williams}, R. N.~M. 2009, in The Magellanic System: Stars, Gas, and Galaxies,
  ed. J.~T. {Van Loon} \& J.~M. {Oliveira}, Vol. 256, 443--453,
  \dodoi{10.1017/S1743921308028846}

\bibitem[{{Yew} {et~al.}(2021){Yew}, {Filipovi{\'c}}, {Stupar}, {Points},
  {Sasaki}, {Maggi}, {Haberl}, {Kavanagh}, {Parker}, {Crawford}, {Vukoti{\'c}},
  {Uro{\v{s}}evi{\'c}}, {Sano}, {Seitenzahl}, {Rowell}, {Leahy}, {Bozzetto},
  {Maitra}, {Leverenz}, {Payne}, {Park}, {Alsaberi}, \& {Pannuti}}]{Yew21}
{Yew}, M., {Filipovi{\'c}}, M.~D., {Stupar}, M., {et~al.} 2021, \mnras, 500,
  2336, \dodoi{10.1093/mnras/staa3382}

\bibitem[{{Zhang} {et~al.}(2014){Zhang}, {Chu}, {Williams}, {Jiang}, {Chen}, \&
  {Gruendl}}]{Zhang2014}
{Zhang}, N.-X., {Chu}, Y.-H., {Williams}, R.~M., {et~al.} 2014, \apj, 792, 58,
  \dodoi{10.1088/0004-637X/792/1/58}

\end{thebibliography}

\end{document}